\documentclass[11pt,a4paper]{article} 
\pdfoutput=1
\usepackage{jheppub,hyperref,mdwlist}
\usepackage{amsmath}
\usepackage{graphicx}
\usepackage{subfigure}
\usepackage{amssymb}
\usepackage{xcolor}
\usepackage{multirow}
\usepackage{cancel}
\usepackage{color}
\usepackage{listings}
\usepackage{verbatim}
\usepackage{float}
\usepackage{slashed}
\usepackage{mathtools}
\usepackage{soul}

\usepackage{tikz}
\usetikzlibrary{arrows,shapes}
\usetikzlibrary{trees}
\usetikzlibrary{matrix,arrows} 				
\usetikzlibrary{positioning}				
\usetikzlibrary{calc,through}				
\usetikzlibrary{decorations.pathreplacing}  
\usepackage{pgffor}							

\usetikzlibrary{decorations.pathmorphing}	
\usetikzlibrary{decorations.markings}
\tikzset{
    vector/.style={decorate, decoration={snake}, draw},
	provector/.style={decorate, decoration={snake,amplitude=2.5pt}, draw},
	antivector/.style={decorate, decoration={snake,amplitude=-2.5pt}, draw},
    fermion/.style={draw=black, postaction={decorate},
        decoration={markings,mark=at position .55 with {\arrow[draw=black]{>}}}},
    fermionbar/.style={draw=black, postaction={decorate},
        decoration={markings,mark=at position .55 with {\arrow[draw=black]{<}}}},
    fermionnoarrow/.style={draw=black},
    gluon/.style={decorate, draw=black,
        decoration={coil,amplitude=4pt, segment length=5pt}},
    scalar/.style={dashed,draw=black, postaction={decorate},
        decoration={markings,mark=at position .55 with {\arrow[draw=black]{>}}}},
    scalarbar/.style={dashed,draw=black, postaction={decorate},
        decoration={markings,mark=at position .55 with {\arrow[draw=black]{<}}}},
    scalarnoarrow/.style={dashed,draw=black},
    electron/.style={draw=black, postaction={decorate},
        decoration={markings,mark=at position .55 with {\arrow[draw=black]{>}}}},
	bigvector/.style={decorate, decoration={snake,amplitude=4pt}, draw},
}

\tikzstyle{block} = [draw, rectangle, 
    minimum height=3em, minimum width=6em]

\usepackage{amsmath}
\usepackage{wrapfig}
\usepackage{cleveref}

\newcommand{\be}{\begin{equation}}
\newcommand{\ee}{\end{equation}}
\newcommand{\beq}{\begin{equation}}
\newcommand{\eeq}{\end{equation}}
\newcommand{\bea}{\begin{eqnarray}}
\newcommand{\eea}{\end{eqnarray}}
\newcommand{\besp}{\begin{equation}\begin{split}}
\newcommand{\eesp}{\end{split}\end{equation}}

\newcommand{\Eq}[1]{Eq.~(\ref{#1})}

\newcommand{\Dfbd}{\mathord{\buildrel{\lower3pt\hbox{$\scriptscriptstyle\leftrightarrow$}}\over {D}_{\mu}}}
\newcommand{\ave}[1]{\left\langle #1\right\rangle}

\def\mL{\mathcal{L}}
\def\mM{\mathcal{M}}

\def\mO{\mathcal{O}}
\def\mP{\mathcal{P}}

\def\veck{\vec{k}}
\def\vecp{\vec{p}}

\def\vec{\mathbf}

\def\0{\textbf{0}}
\def\1{\textbf{1}}
\def\2{\textbf{2}}
\def\3{\textbf{3}}
\def\4{\textbf{4}}
\def\5{\textbf{5}}
\def\6{\textbf{6}}
\def\7{\textbf{7}}
\def\8{\textbf{8}}
\def\9{\textbf{9}}

\def\d{\text{d}}
\def\hc{\text{h.c.}}

\def\x{\textbf{x}}
\def\p{\textbf{p}}

\definecolor{RoyalBlue}{cmyk}{1, 0.50, 0, 0}

\begin{document}

\title{When inverse seesaw meets inverse electroweak phase transition: a novel path to leptogenesis}

\author[a,b,c]{Wen-Yuan Ai}
\affiliation[a]{State Key Laboratory of Dark Matter Physics,\\
Tsung-Dao Lee Institute and School of Physics and Astronomy,
Shanghai Jiao Tong University, Shanghai 201210, China}
\affiliation[b]{
Key Laboratory for Particle Astrophysics and Cosmology (MOE)
\& Shanghai Key Laboratory for Particle Physics and Cosmology}
\affiliation[c]{Marietta Blau Institute for Particle Physics, Austrian Academy of Sciences,\\ Dominikanerbastei 16, A-1010 Vienna, Austria}

\author[d]{Peisi Huang}
\affiliation[d]{Department of Physics and Astronomy, University of Nebraska, Lincoln, NE 68588, USA}

\author[e]{and Ke-Pan Xie}
\affiliation[e]{School of Physics, Beihang University, Beijing 100191, China}

\emailAdd{wenyuan.ai@oeaw.ac.at}
\emailAdd{peisi.huang@unl.edu}
\emailAdd{kpxie@buaa.edu.cn}

\abstract{
We propose a new nonthermal leptogenesis mechanism triggered by the cosmic first-order phase transition. The Standard Model is extended with two generations of TeV-scale vectorlike leptons. The lighter generation gives rise to an {\it inverse} electroweak phase transition of the Higgs field at $T\sim200~{\rm GeV}$, restoring the symmetry, and resulting in relativistic bubble expansion in the space. The heavier generation is responsible for neutrino masses via the {\it inverse} seesaw mechanism. The interaction between bubble walls and particles in the plasma abundantly produces the vectorlike leptons, and they subsequently undergo CP-violating decay to generate the baryon asymmetry. This mechanism is testable at current and future particle experiments.
}

\maketitle
\flushbottom

\section{Introduction}

The observed baryon asymmetry of the Universe (BAU), $Y_B=n_B/s\approx8.58\times10^{-11}$~\cite{ParticleDataGroup:2024cfk}, is a mystery in particle physics and cosmology that calls for physics beyond the Standard Model (SM). A possible solution is leptogenesis, which first generates a lepton asymmetry, and then partly converts it to the BAU via the electroweak (EW) sphaleron process~\cite{Kuzmin:1985mm}. The most studied scenario is the out-of-equilibrium and CP-violating decay of heavy right-handed neutrinos (RHNs), which can be embedded into the seesaw mechanism~\cite{Minkowski:1977sc}, serving as an elegant explanation to the BAU and SM neutrino mass origin~\cite{Buchmuller:2004nz,Buchmuller:2005eh,Davidson:2008bu,Pilaftsis:2009pk,Xing:2020ald}. Depending on the initial condition, leptogenesis can be classified into thermal and nonthermal types. In the former case, RHNs are originally in the thermal bath, but as the Universe expands they depart from equilibrium and subsequently decay~\cite{Fukugita:1986hr,Luty:1992un}. In the latter case, RHNs are absent in the plasma until they are produced via nonthermal processes at a temperature much lower than their mass, and then decay rapidly to create the BAU. The traditional scenario was the inflaton decay during reheating or preheating~\cite{Lazarides:1990huy,Asaka:1999jb,Giudice:1999fb}, while in recent years it has been noticed that cosmic first-order phase transitions (FOPTs) can also efficiently produce RHNs and realize nonthermal leptogenesis.

A FOPT is the Universe transition from a false vacuum to a true vacuum, via bubble nucleation, expansion, and merging, typically driven by a scalar potential with two minima separated by a barrier~\cite{Mazumdar:2018dfl,Caprini:2019egz,Athron:2023xlk}. If it occurs at a temperature much lower than the scale of the scalar, it is referred to as a supercooled transition. A supercooled FOPT is usually accompanied by bubbles expanding at relativistic velocities, enabling production of RHNs much heavier than the FOPT temperature. For example, if the RHNs acquire mass via the scalar field, the expanding bubble sweeps the abundant massless RHNs into its interior, where they undergo a sudden mass increase and decay, known as the mass-gain mechanism~\cite{Baldes:2021vyz,Huang:2022vkf,Chun:2023ezg,Dichtl:2023xqd}. Another possibility is RHN production from the scattering between the bubble walls and light particles in the plasma, which are highly boosted in the wall-frame and hence very energetic~\cite{Azatov:2021irb}. Besides, bubble collisions can be a direct source of RHN production~\cite{Cataldi:2024pgt}.\footnote{Here we focus on bubble-assisted leptogenesis via RHN decay. For mechanisms via neutrino transport in the vicinity of bubble walls, see Refs.~\cite{Pascoli:2016gkf,Long:2017rdo,Fernandez-Martinez:2020szk}.} Compared with thermal leptogenesis, bubble-assisted leptogenesis does not suffer from thermal washout, expands the viable parameter space, and leads to extra signals such as stochastic gravitational waves (GWs).

\begin{figure}
    \centering
    \includegraphics[width=0.8\linewidth]{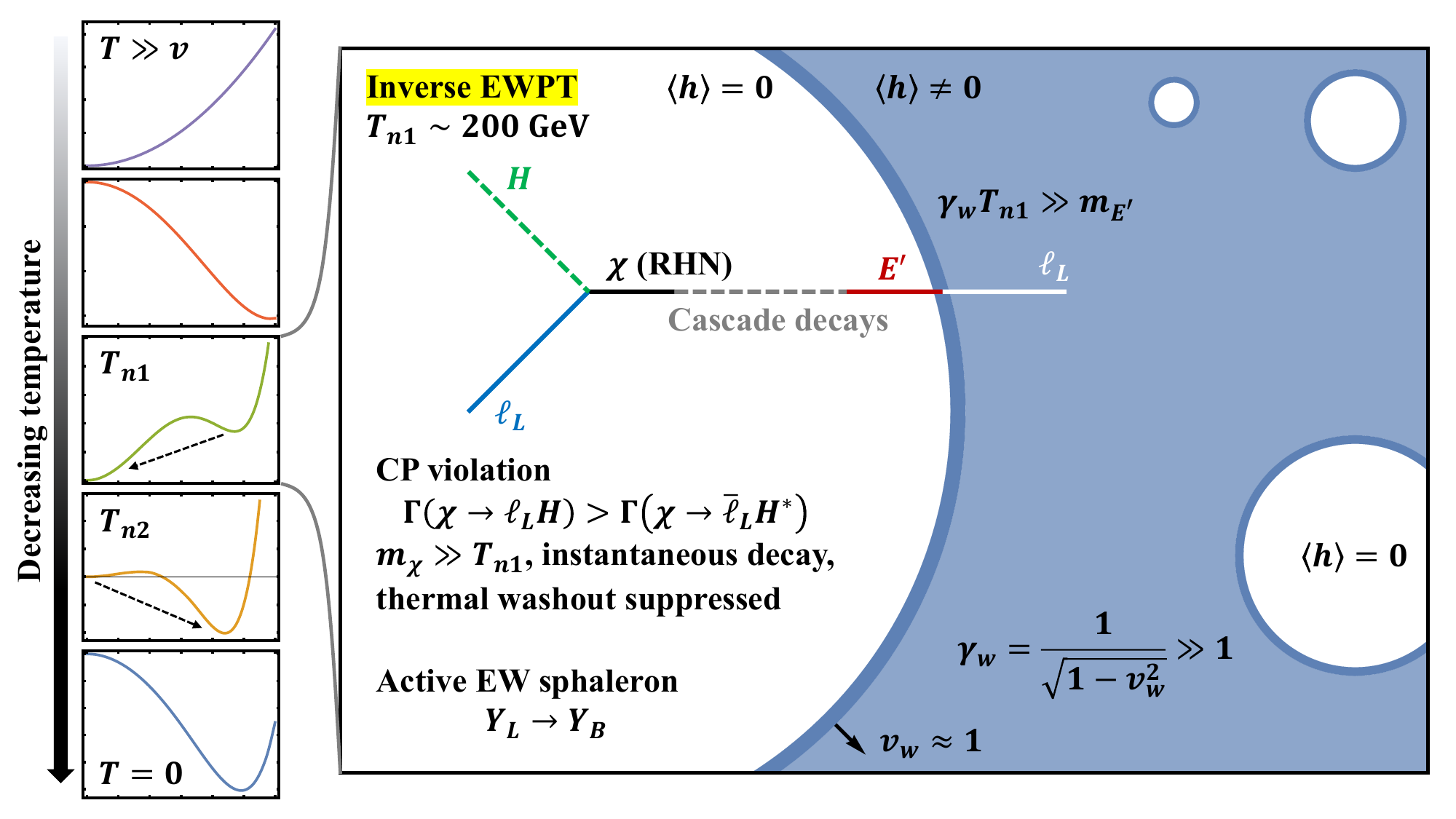}
    \caption{Sketch of our mechanism. When the Universe cools down, the Higgs effective potential evolves and triggers two successive first-order EWPTs, first an inverse one and then a direct one. During the inverse EWPT, RHNs are produced via the scattering between SM light leptons $\ell_L$ and the bubble walls, and then decay immediately and generate the BAU. See the text for details.}
    \label{fig:sketch}
\end{figure}

In this work, we propose a novel mechanism based on a first-order electroweak phase transition (EWPT) of the SM Higgs field $h$. At first glance, bubble-assisted leptogenesis is incompatible with EWPT: the conventional {\it direct} EWPT proceeds from $\ave{h}=0$ to $\ave{h}\neq0$, thus the bubble interior is the EW-breaking vacuum, where the sphaleron is inefficient, preventing the conversion of lepton asymmetry generated from RHN decay into the BAU. However, we will show that leptogenesis can occur during an {\it inverse} EWPT, where the transition is from $\ave{h}\neq0$ to $\ave{h}=0$, and the bubble interior is the EW-symmetric vacuum where the EW sphaleron is active. The inverse EWPT occurs at $T_{n1}\sim200$ GeV, and it is not supercooled; however, it features relativistic bubble expansion driven by the loss of SM particle masses as they enter the bubbles. RHNs with mass $m_\chi\gg T_{n1}$ are produced by the scattering between SM left-handed leptons $\ell_L$ and the bubble walls, and then immediately decay to generate the BAU. The mechanism is sketched in Fig.~\ref{fig:sketch}, with three key points summarized below.

First, the nature of EWPT. The SM EWPT is a smooth crossover without bubble nucleation. Various new physics effects can induce a barrier in the Higgs potential, resulting in a first-order EWPT~\cite{Chung:2012vg}. But typically the transition is a direct one, tunneling from $\ave{h}=0$ to $\ave{h}\neq0$. What we need is an unusual inverse EWPT from $\ave{h}\neq0$ to $\ave{h}=0$, and this can be realized by extending the SM with one generation of vectorlike leptons (VLLs)~\cite{Angelescu:2018dkk}.\footnote{Note our inverse EWPT occurs in a cooling Universe, different from inverse FOPTs during reheating when the temperature is increasing~\cite{Buen-Abad:2023hex,Barni:2025ced,Ai:2024cka,Dent:2024bhi}. Recently, inverse FOPT during the cosmic cooling is also studied in a supersymmetric model~\cite{Barni:2025mud}.} In its appropriate parameter space, the evolution of the Higgs potential is shown on the left of Fig.~\ref{fig:sketch}, providing the necessary environment for our mechanism.

Second, the origin of neutrino mass and leptogenesis. We find that a single generation of VLLs can be embedded into the inverse seesaw mechanism~\cite{Wyler:1982dd,Mohapatra:1986aw,Mohapatra:1986bd} and explain the neutrino oscillation data. In this setup, the neutral VLLs play the role of RHNs, and the seesaw interactions also yield the requisite decay vertices for leptogenesis. Although only one generation of VLLs can realize both the inverse EWPT and the inverse seesaw, it is insufficient for successful leptogenesis. This limitation arises because suppressing thermal washout demands RHN mass of $\sim4~{\rm TeV}$ (or equivalently $m_\chi\gtrsim20\,T_{n1}$, which we derive later), whereas the inverse EWPT favors VLL mass at $\sim{\rm TeV}$. To reconcile these, we introduce two separate VLL generations: one for EWPT and the other for leptogenesis.

Third, RHN production. To explain the observed BAU, a sufficient abundance of RHNs must be generated during the inverse EWPT. The simplest approach employs the seesaw vertices to generate RHNs via $\ell_L$ scattering with the bubble walls. However, this turns out to be inadequate because the magnitude of the couplings is fixed by SM neutrino mass, which is too small for TeV-scale RHNs. Consequently, we use the coupling between the charged singlet VLL and $\ell_L$, which is less constrained experimentally and can be relatively sizable. This interaction efficiently produces charged VLLs, which cascade‌ decay into RHNs, ultimately generating the lepton asymmetry necessary for leptogenesis.

This paper is organized as follows. In section~\ref{sec:model}, we introduce the model, demonstrating that an inverse EWPT can be triggered by the first generation of VLLs and the inverse seesaw mechanism can be realized by the second generation of VLLs. Then we detail the bubble-assisted leptogenesis in section~\ref{sec:leptogenesis}, describing the calculation formalism and showing the numerical results. Finally, the conclusion is given in section~\ref{sec:conclusion}.

\section{The model}\label{sec:model}

\subsection{The first-generation VLLs: inverse EWPT}\label{subsec:ewpt}

VLLs are color singlet Dirac fermions with the following $SU(2)_L\times U(1)_Y$ quantum numbers:
\be
F=\begin{pmatrix} F^0 \\ F^-\end{pmatrix}\sim \2_{-1/2}\,,\quad N\sim \1_{0}\,,\quad E\sim \1_{-1}\,.
\ee
In other words, $F$ and $E$ have the same gauge quantum number with the SM $\ell_L$ and $e_R$, respectively, while $N$ is a gauge singlet. The Lagrangian reads
\begin{multline}\label{LVLL}
\mL_{\rm VLL}=\bar F(i\slashed{D}-m_F)F +\bar N(i\slashed{\partial}-m_N)N +\bar E(i\slashed{D}-m_E)E\\
-y_{N_L}\bar F_R \tilde H N_L-y_{N_R}\bar F_L\tilde H N_R-y_{E_L}\bar F_RH E_L-y_{E_R}\bar F_LH E_R+\hc\,,
\end{multline}
where $H=\left(G^+,(h+iG^0)/\sqrt{2}\right)^T$ is the SM Higgs doublet and $\tilde{H}\equiv i\sigma^2 H^*$ is its charge conjugation, and $D_\mu$ denotes the SM gauge covariant derivative.

The bare masses for $F$, $N$, and $E$ are respectively $m_F$, $m_N$, and $m_E$, and the Yukawa couplings induce mixings between the VLLs after the EW symmetry breaking. Specifically, the electric neutral VLLs $(F^0, N)$ and charged VLLs $(F^-,E)$ form two sets of mass terms
\be
-\begin{pmatrix}\bar F^0 & \bar N\end{pmatrix}_L\begin{pmatrix}m_F &\frac{y_{N_R}v}{\sqrt{2}} \\ \frac{y_{N_L}v}{\sqrt{2}} & m_N\end{pmatrix}\begin{pmatrix}F^0 \\ N\end{pmatrix}_R-\begin{pmatrix}\bar F^- & \bar E\end{pmatrix}_L\begin{pmatrix}m_F &\frac{y_{E_R}v}{\sqrt{2}} \\ \frac{y_{E_L}v}{\sqrt{2}} & m_E\end{pmatrix}\begin{pmatrix}F^- \\ E\end{pmatrix}_R+\hc\,,
\ee
with $v=246$ GeV being the Higgs vacuum expectation value (VEV) at zero temperature. Diagonalizing the mass matrices, we then obtain mass eigenstates $(N_1, N_2)$ and $(E_1,E_2)$, with mass squared eigenvalues
\begin{multline}\label{MX12}
m_{X_{1,2}}^2=\frac{m_F^2+m_X^2}{2}+\frac{y_{X_L}^2+y_{X_R}^2}{4}v^2\\
\mp\frac12\sqrt{\left(m_F^2+m_X^2+\frac{y_{X_L}^2+y_{X_R}^2}{2}v^2\right)^2-\left(2m_F m_X-y_{X_L}y_{X_R}v^2\right)^2}\,,
\end{multline}
where $X=N$ or $E$.

We now examine the finite-temperature behavior of the model. The tree-level Higgs potential reads $\mu^2|H|^2+\lambda|H|^4$, or in the unitary gauge,
\be\label{V0}
V_0(h)=\frac{\mu^2}{2}h^2+\frac{\lambda}{4}h^4\equiv-\frac{m_h^2}{4}h^2+\frac{m_h^2}{8v^2}h^4\,,
\ee
where $m_h=125$ GeV is the Higgs boson mass. The EW symmetry is spontaneously broken, since $V_0(h)$ is minimized at $h=v$. In the early Universe, the thermal corrections lead to a temperature-dependent effective potential $V_T(h)$. Typically, the corrections are dominated by light particles whose mass originates from the Higgs VEV, while contributions from heavy particles with masses beyond the SM (BSM) are Boltzmann-suppressed. As a result, $V_T(h)$ generally exhibits a form of $(\mu^2+c_2T^2)h^2/2-c_3Th^3/3+\lambda h^4/4$, where $c_2$ and $c_3$ are positive coefficients determined by the interactions between Higgs and light particles~\cite{Chung:2012vg}. When $T\gg v$, the $c_2$ term dominates $V_T(h)$, restoring the EW symmetry; when $T$ drops to $\mO(100~{\rm GeV})$, the interplay between the $c_2$ and $c_3$ terms yields a barrier for $V_T(h)$, triggering a direct first-order EWPT. However, our model differs from this conventional picture significantly.

VLLs affect $V_T(h)$ through the field-dependent squared mass, denoted by $\mM_{X_{1,2}}^2(h)$, which can be obtained by replacing $v$ with the background field $h$ in \Eq{MX12}. As an illustration, we only consider the neutral VLLs, and set $m_F=m_N=m$ and $y_{N_L}=y_{N_R}=\sqrt{2}\tilde y$ for simplicity, then $\mM_{N_{1,2}}^2(h)=(m\pm\tilde yh)^2$. If both $m$ and $\tilde yh$ are of $\mO({\rm TeV})$, while $|m-\tilde y h|\sim100~{\rm GeV}$, then at $T\sim100$ GeV, the contribution from the heavier state $N_2$ is suppressed by the Boltzmann factor $e^{-(m+\tilde yh)/T}$. Conversely, the lighter state $N_1$ contributes a term $\sim T^2(m-\tilde yh)^2/2$ to $V_T(h)$. This term is minimized at $h\neq0$, preventing the symmetry restoration. When combined with the $c_2$ and $c_3$ terms, this can lead to the inverse EWPT. The nonvanishing bare Dirac mass $m$ is crucial for inducing an inverse FOPT, and this is why vectorlike BSM fermions are required.

\begin{figure}
    \centering
    \includegraphics[scale=0.55]{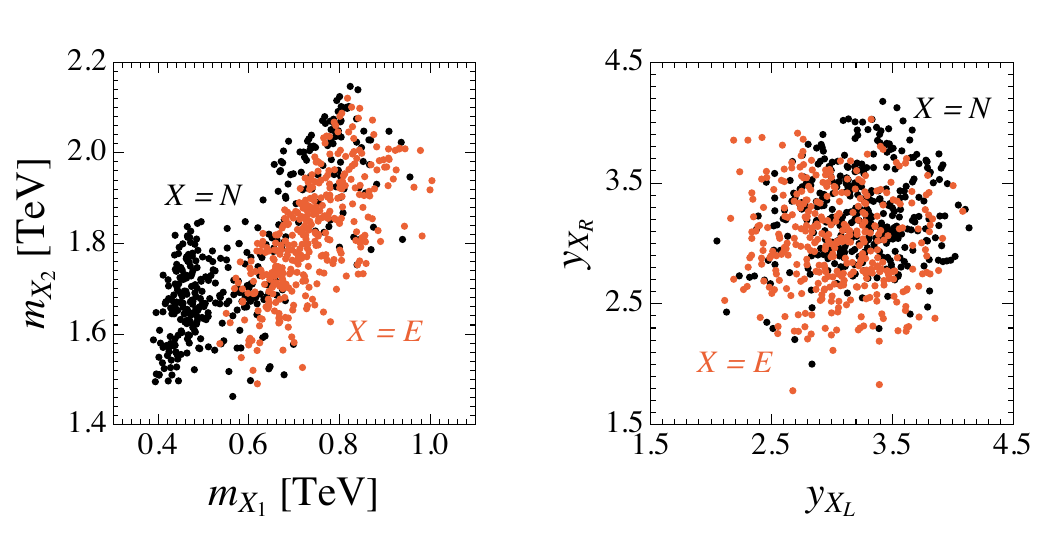}
    \caption{The inverse EWPT parameter space projected to two-dimensional plots. Left: the physical masses $m_{X_1}$ v.s. $m_{X_2}$. Right: the Yukawa couplings $y_{X_L}$ v.s. $y_{X_R}$. The black and dark red colors represent $X=N$ and $E$, respectively.}
    \label{fig:EWPT_parameter_space}
\end{figure}

The above simplified analysis indicates that TeV-scale VLLs with sizable $y_{N_{L,R}}$ and $y_{E_{L,R}}$ may realize the inverse EWPT. We have confirmed this under the full one-loop thermal effective potential, with detailed expressions and the treatment of EWPT dynamics provided in Appendix~\ref{app:ewpt}, following standard finite-temperature field theory. The numerical results are presented here. We scan over
\be
m_F,\, m_N,\, m_E\in[0.8~{\rm TeV},\,1.5~{\rm TeV}]\,;\quad y_{N_{L,R}}, \, y_{E_{L,R}}\in[0.1,\,\sqrt{8\pi}]\,,
\ee
requiring $y_{N_L}^2+y_{E_R}^2<8\pi$ and $y_{N_R}^2+y_{E_L}^2<8\pi$ due to the unitarity constraint~\cite{Angelescu:2018dkk}. To be consistent with the current Higgs and EW measurements, we require the Higgs di-photon signal strength $\mu_{\gamma\gamma}=\Gamma(h\to\gamma\gamma)/\Gamma_{\rm SM}(h\to\gamma\gamma)$ and the oblique parameters~\cite{Peskin:1990zt,Peskin:1991sw} to be within the $2\sigma$ range of the Particle Data Group results~\cite{ParticleDataGroup:2024cfk},\footnote{The calculation formulas of $\mu_{\gamma\gamma}$ and oblique parameters are taken from Refs.~\cite{Kearney:2012zi} and \cite{Joglekar:2012vc}, respectively.} namely $\mu_{\gamma\gamma}=1.10\pm0.12$, $S=-0.04\pm0.2$, and $T=0.01\pm0.24$. The parameter space that can realize the thermal history depicted in Fig.~\ref{fig:sketch} is shown in Fig.~\ref{fig:EWPT_parameter_space}. Large Yukawa couplings are necessary for an inverse EWPT, and three additional comments are in order. First, the Landau pole appears above $\mO(10~{\rm TeV})$. Second, such large couplings are the reason of introducing a full generation of VLLs (i.e., $F$, $N$, and $E$) since they respect the custodial symmetry and hence protect the $T$-parameter. Third, as demonstrated in other models -- such as the real-singlet extended SM~\cite{Niemi:2024vzw,Ramsey-Musolf:2024ykk}, the inert doublet model~\cite{Laine:2017hdk}, and the Abelian Higgs model~\cite{Funakubo:2012qc} -- two-loop contributions to the finite-temperature effective potential~\cite{Ekstedt:2022bff} can moderately or even significantly alter the phase transition strength and critical temperature. A similar refinement may therefore be necessary in the present VLL model and change the shape of the viable parameter space. More discussions on EWPTs in the VLL model can be found in Ref.~\cite{Angelescu:2018dkk}.

\begin{figure}
    \centering
    \includegraphics[scale=0.75]{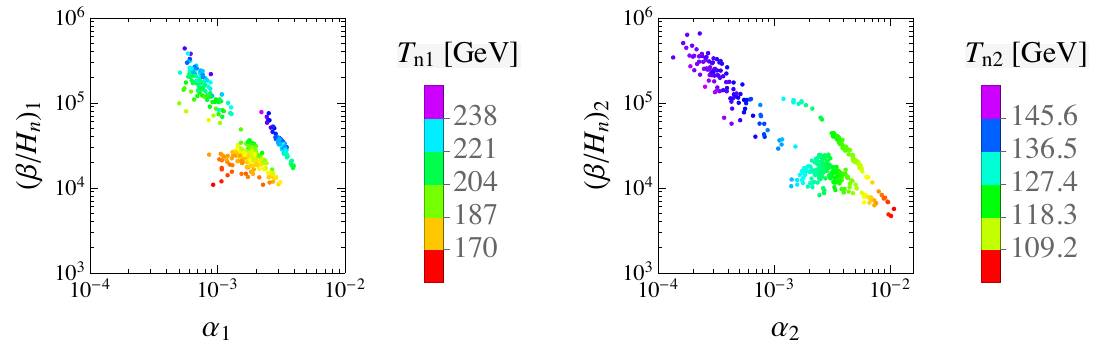}
    \caption{The phase transition parameters for the inverse (left) and direct (right) EWPTs.}
    \label{fig:EWPT_parameters}
\end{figure}

The first (and inverse) EWPT is $h$: $v_{n1}\to0$ at $T_{n1}$, while the second (and direct) EWPT is $h$: $0\to v_{n2}$ at $T_{n2}$. For each EWPT, we define the ratio of latent heat to the radiation energy density as
\be
\alpha=\frac{1}{\pi^2g_*T_n^4/30}\left(\Delta V_T-\frac{T}{4}\frac{\partial\Delta V_T}{\partial T}\right)\Big|_{T_n},
\ee
with $\Delta V_T$ being the positive effective potential difference between the true and false vacua, and $g_*$ being the number of effective degrees of freedom (d.o.f.) in the plasma. Note that with this definition, the $\alpha$ is positive for the inverse EWPT. According to Ref.~\cite{Barni:2025mud}, this phase transition does not necessarily give inverse hydrodynamics. We further define $\beta/H_n$ as the ratio of the Hubble time scale to EWPT duration. The parameters of the two EWPTs are shown in Fig.~\ref{fig:EWPT_parameters}. The inverse EWPT typically occurs at $\sim200$ GeV, while the subsequent direct EWPT occurs at $\sim100$ GeV.

Although $T_{n1}$ is around the EW scale, implying that the inverse EWPT is not supercooled, relativistic bubble expansion still occurs. In general, bubble wall velocity $v_w$ is determined by the balance of the outward pressure and the inward friction. This usually involves hydrodynamics (see e.g.~\cite{Laine:1993ey,Kurki-Suonio:1995rrv,Espinosa:2010hh}) and the complicated interaction between the plasma and the bubble wall that calls for solving the integro-differential Boltzmann equations (see e.g.~\cite{Liu:1992tn,Moore:1995ua, Moore:1995si,DeCurtis:2022hlx,Laurent:2022jrs,Ekstedt:2024fyq,Branchina:2025jou}). However, in the relativistic limit, the analysis becomes simple, and one needs only compare with the vacuum pressure and particle transition pressure, including the particle mass-gain $\mP_{1\to1}$ and particle-splitting $\mP_{1\to N}$ contributions~\cite{Bodeker:2009qy,Bodeker:2017cim,Azatov:2020ufh,Hoche:2020ysm,Gouttenoire:2021kjv,Ai:2023suz,Azatov:2023xem}. Note that the false and true vacua of an EWPT are defined by the finite-temperature effective potential $V_T$, whose difference $\Delta V_T$ is always positive; while the vacuum pressure is the {\it zero-temperature} potential difference $\Delta V$, which may be positive or negative. In a direct EWPT, $\Delta V>0$, and $v_w$ is obtained from $\Delta V=\mP_{1\to1}+\mP_{1\to N}$. Consequently, relativistic bubble walls are available only when $\Delta V$ is large enough, which demands a supercooling. In contrast, during an inverse EWPT, $\Delta V<0$ and $\mathcal{P}_{1\to 1}$ is also negative due to SM particles losing masses when entering the bubble. This means that $\Delta V$ and $\mathcal{P}_{1\to 1}$ change their roles as driving force and friction~\cite{Buen-Abad:2023hex}, and $v_w$ is given by $|\mP_{1\to1}| = |\Delta V|+ \mP_{1\to N}$, which yields $v_w\approx1$.\footnote{We note however that, there is a friction caused purely from hydrodynamic effects~\cite{Konstandin:2010dm} which induces a pressure barrier at the Jouguet velocity~\cite{Cline:2021iff,Laurent:2022jrs,Ai:2023see}. Hence, one in principle has to confirm that this hydrodynamic obstruction cannot stop the wall from entering the relativistic regime~\cite{Ai:2024shx}.} Based on the $1\to2$ splitting result~\cite{Azatov:2024auq}, we obtain $\gamma_w=1/\sqrt{1-v_w^2} \sim\mO(10^2)$.

\subsection{The second-generation VLLs: inverse seesaw}

The second generation of VLLs are labeled by ``\,$\prime$\,'', i.e., $F'$, $N'$, and $E'$. They have the same quantum numbers with the first generation ones but with heavier masses. Similar to \Eq{LVLL}, we can write down the VLL Lagrangian including the (gauge) kinetic terms, bare mass terms, and Higgs portal couplings, namely
\begin{multline}\label{LVLL2}
\mL'_{\rm VLL}=\bar F'(i\slashed{D}-m'_F)F' +\bar N'(i\slashed{\partial}-m'_N)N' +\bar E'(i\slashed{D}-m_E')E'\\
-y'_{N_L}\bar F'_R \tilde H N'_L-y'_{N_R}\bar F'_L\tilde H N'_R-y'_{E_L}\bar F'_RH E'_L-y'_{E_R}\bar F'_LH E'_R+\hc
\end{multline}
For neutrino mass and leptogenesis, we further add
\begin{multline}\label{LVLLp2}
\mL'_{\nu\text{-VLL}}=
-\frac{\mu_N}{2}\bar N'_LN'^c_L-\lambda_N\bar F'_L\tilde HN'^c_L\\
-\sum_\alpha y_D^\alpha\bar\ell_L^\alpha\tilde HN'_R-\sum_\alpha\mu_D^\alpha\bar\ell_L^\alpha F'_R-\sum_\alpha \lambda_D^\alpha\bar\ell_L^\alpha HE'_R+\hc\,,
\end{multline}
where $\alpha=e$, $\mu$, $\tau$, representing the SM lepton flavors, and $N'^{c}_L=i\gamma^2N'^*_L$ is the charge conjugation under the Weyl representation. The first line breaks $U(1)_{B-L}$, while the second line describes the portal interactions between SM leptons and VLLs, where we only include the $\ell_L$-terms and neglect the $e_R$-terms for simplicity. To realize the nonthermal leptogenesis, we assume
\be\label{mass_hierarchy}
m'_E>m'_F>m'_N\sim4~{\rm TeV},\quad y'_{N_{L,R}},\,y'_{E_{L,R}}\sim1\,.
\ee
The second-generation VLLs are heavier than the first-generation ones, while the couplings to the SM Higgs and leptons are much weaker, and hence their influence on phase transition dynamics is negligible.

After the EW symmetry breaking, the neutral part of VLL model can be nicely embedded into the neutrino inverse seesaw model~\cite{Wyler:1982dd,Mohapatra:1986aw,Mohapatra:1986bd}. Let $h=v$, the neutral fermion mass terms from \Eq{LVLL2} and \Eq{LVLLp2} can be reorganized as
\be\label{Npmass}
\mL_{\rm mass}^{h\to v}=-\sum_{\alpha,i}m_D^{\alpha i}\bar\nu_L^\alpha\nu_R^i-\sum_{i,j}M_R^{ij}\bar\nu_R^{i,c}S_L^{j,c}-\sum_{i,j}\frac12\mu_{ij}\bar S_L^{i}S_L^{j,c}+\hc\,,
\ee
with $i=1$, 2 being the heavy neutrino indices, and 
\be
\nu_R^1=N'_R\,,\quad \nu_R^2=F'^0_R\,,\quad  S_L^1=N'_L\,,\quad S_L^2=F'^0_L\,.
\ee
The mass matrices are
\be
m_D^{\alpha i}=\begin{pmatrix}\frac{y_D^ev}{\sqrt{2}} & \mu_D^e \\ \frac{y_D^\mu v}{\sqrt{2}} & \mu_D^\mu\\ \frac{y_D^\tau v}{\sqrt{2}} & \mu_D^\tau\end{pmatrix},\quad
M_R^{ij}=\begin{pmatrix}m'_N & \frac{y'_{N_R}v}{\sqrt{2}} \\ \frac{y'^*_{N_L}v}{\sqrt{2}} & m'_F\end{pmatrix},\quad
\mu_{ij}=\begin{pmatrix}\mu_N & \frac{\lambda_Nv}{\sqrt{2}} \\ \frac{\lambda_Nv}{\sqrt{2}} & 0\end{pmatrix}.
\ee
Given these matrices, the mass terms in \Eq{Npmass} can be written neatly as
\be
\mL_{\rm mass}^{h\to v}=-\frac12\begin{pmatrix}\bar\nu_L & \bar\nu_R^c & \bar S_L\end{pmatrix}M_\nu\begin{pmatrix}\nu_L^c\\\nu_R\\ S_L^c\end{pmatrix}+\hc\,,\quad M_\nu=\begin{pmatrix}\textbf{0}_{3\times 3} & m_D & \textbf{0}_{3\times 2}\\ m_D^T & \textbf{0}_{2\times2} & M_R \\ \textbf{0}_{2\times 3} & M_R^T & \mu\end{pmatrix}.
\ee
This is the form of the inverse seesaw mechanism, or more precisely the minimal inverse seesaw~\cite{Malinsky:2009df,Abada:2014vea} because only two sets of heavy neutrinos are introduced, and the SM lightest neutrino is massless.

Given $\mu$, $m_D\ll M_R$, we can define the unitary transformation
\be
U_\nu=\begin{pmatrix}\textbf{1}_{3\times3} &-m_D^*(M_R^\dagger)^{-1}\mu^\dagger(M_R^*)^{-1} & m_D^*(M_R^\dagger)^{-1} \\ (M_R^T)^{-1}\mu M_R^{-1}m_D^T & \textbf{1}_{2\times2} & \textbf{0}_{2\times2} \\ -M_R^{-1}m_D^T & \textbf{0}_{2\times2} & \textbf{1}_{2\times2}\end{pmatrix}
\ee
which satisfies $U_\nu^\dagger U_\nu=\textbf{1}_{7\times7}+\mO(m_D^2/M_R^2)$ to block diagonalize \Eq{Npmass} as
\be\label{block}
U_\nu^TM_\nu U_\nu=\left(\begin{array}{c|cc}m_D(M_R^T)^{-1}\mu M_R^{-1}m_D^T &  &  \\ \hline  & \textbf{0}_{2\times2} & M_R \\
 & M_R^T  & \mu \end{array}\right)+\mO(m_D^2/M_R^2)\,.
\ee
The bottom-right block is the heavy neutrino mass, and it gives two pairs of near-degenerate Majorana fermions coming from $N'$ and $F'^0$, respectively. The top-left block is the SM neutrino mass term
\be
\mL_{\nu\text{-mass}}^{h\to v}=-\frac12\bar\nu_Lm_\nu\nu_L^c-\frac12\bar\nu_L^cm_\nu^\dagger \nu_L\,,
\ee
with $m_\nu=m_D(M_R^T)^{-1}\mu M_R^{-1}m_D^T$. We can see that the SM neutrino mass has a double expression factor $\sim \mu(m_D^2/M_R^2)$, different from the conventional type-I seesaw ($\sim m_D^2/M_R$), allowing for larger $y_D$ couplings. Although the treatment here follows the standard inverse seesaw technique, we should keep in mind that $\nu_R^2$ and $S_L^2$ here are {\it not} SM gauge singlets, as they come from the neutral component of a weak doublet $F'$.

We adopt the Casas-Ibarra (CI) scheme~\cite{Casas:2001sr} to parametrize the $m_D$ matrix using the method in Ref.~\cite{Dolan:2018qpy}. Since $M_R$ has rank 2, one generation of the SM neutrinos should be massless. The unitary Pontecorvo-Maki-Nakagawa-Sakata (PMNS) matrix diagonalizes $m_\nu^\dagger$ via $U_{\rm PMNS}^Tm_\nu^\dagger U_{\rm PMNS}=m_n\equiv{\rm diag}\{0,m_2,m_3\}$, where we have assumed the normal ordering of neutrino mass and hence $m_1=0$. $U_{\rm PMNS}$ can be parametrized as
\be
\begin{pmatrix} c_{12}c_{13} & s_{12}c_{13} & s_{13}e^{-i\delta_{\rm CP}} \\
-s_{12}c_{23}-c_{12}s_{23}s_{13}e^{i\delta_{\rm CP}} & c_{12}c_{23}-s_{12}s_{23}s_{13}e^{i\delta_{\rm CP}} & s_{23}c_{13} \\
s_{12}s_{23}-c_{12}c_{23}s_{13}e^{i\delta_{\rm CP}} & -c_{12}s_{23}-s_{12}c_{23}s_{13}e^{i\delta_{\rm CP}} & c_{23}c_{13}\end{pmatrix}\begin{pmatrix}1&0&0\\ 0 & e^{i\frac{\alpha_{21}}{2}}& 0 \\ 0&0& 1\end{pmatrix},
\ee
where $c_{\alpha\beta}\equiv\cos\theta_{\alpha\beta}$ and $s_{\alpha\beta}\equiv\sin\theta_{\alpha\beta}$ with $\theta_{12,23,13}$ being the mixing angles, and there are two CP-violating phases, including the Dirac $\delta_{\rm CP}$ and the Majorana $\alpha_{21}$. We adopt the best fit results of mixing angles and $\delta_{\rm CP}$ from Ref.~\cite{Capozzi:2021fjo}, and set $\alpha_{21}=0$.

Define a $3\times2$ matrix $R=m_n^{-1/2}U_{\rm PMNS}^\dagger m_D(M_R^T)^{-1}\mu^{1/2}$, where
\be\label{mu12}
\mu^{1/2}=\frac{1}{\sqrt{\mu_N+i \sqrt{2}\lambda_{N}v}}\begin{pmatrix}\mu_N+ \frac{i\lambda_{N}v}{\sqrt{2}} & \frac{\lambda_{N}v}{\sqrt{2}} \\
 \frac{\lambda_{N}v}{\sqrt{2}} & \frac{i\lambda_{N}v}{\sqrt{2}}\end{pmatrix}
\ee
with $\mu^{1/2}\mu^{1/2}=\mu$, and $m_n^{-1/2}={\rm diag}\{0,m_2^{-1/2},m_3^{-1/2}\}$. Then the $m_\nu^\dagger$ diagonalization formula can be rewritten as $R^TR=\textbf{1}_{2\times2}$ and $RR^T={\rm diag}\{0,1,1\}$. This means that $R$ can be generally expressed as
\be
R=\begin{pmatrix} 0 & 0 \\ \cos\theta_R & -\sin\theta_R \\ \sin\theta_R & \cos\theta_R\end{pmatrix},
\ee
where $\theta_R$ is a complex number. Now, the Dirac mass term is written as
\be\label{CI}
m_D=U_{\rm PMNS}m_n^{1/2}R(\mu^{1/2})^{-1}M_R^T\,,
\ee
which is called the CI parametrization. Via such a formalism, we can obtain $y_D$ and $\mu_D$ by inputing $U_{\rm PMNS}$, $m_n$, $R$, $\mu$, and $M_R$, and the neutrino oscillation results can always be reproduced. In this work, we assume $y'_{N_{L,R}}$, $y'_{E_{L,R}}$, $\mu_N$, $\lambda_N$, and $\theta_R$ are all real numbers, while $y_D$ and $\mu_D$ can be complex due to the CI parametrization via the imaginary numbers from $\mu^{1/2}$ and $U_{\rm PMNS}$. The CP features of this model will be discussed in section~\ref{subsec:leptogenesis}.

\section{Leptogenesis}\label{sec:leptogenesis}

\subsection{VLL production durning bubble expansion}\label{app:wall_produced}

As outlined in the Introduction, our leptogenesis mechanism relies on nonthermal production of heavy RHNs from bubble expansion. Due to their TeV-scale masses, the VLLs' equilibrium number densities are Boltzmann-suppressed during the inverse EWPT. Nonetheless, due to the lepton portal interactions in the second line of \Eq{LVLLp2}, the SM leptons can experience light-to-heavy transitions~\cite{Azatov:2020ufh,Azatov:2021ifm,Baldes:2022oev,Azatov:2024crd,Ai:2024ikj,Ai:2025bjw,Ramsey-Musolf:2025jyk}, that is $\ell_L\to F'_R$ (via $\mu_D$), $\ell_L\to N'_R$ (via $y_D$), and $\ell_L\to E'_R$ (via $\lambda_D$), when they cross the bubble walls. Such transitions seem impossible since the $\ell_L$ kinetic energy is $\mO(T_{n1})\ll m'_{F,N,E}$. However, in the wall frame $\ell_L$ is boosted to have an energy of $\mO(\gamma_wT_{n1})\gg m'_{F,N,E}$ as $\gamma_w\sim\mO(10^2)$. Besides, the wall provides a space-dependent background field $h(\x)$ that violates the translation invariance that leads to the momentum non-conservation. These two features allow the $\ell_L$ transition to the heavy VLLs on the bubble walls. 

Since $N'$ and $F'$ participate in the inverse seesaw mechanism, the magnitudes of $y_D$ and $\mu_D$ are related to the SM neutrino mass. For TeV-scale $N'$ and $F'$, the relevant couplings are too small to produce considerable VLLs. On the other hand, the $\lambda_D^\alpha $ couplings with $E'_R$ are less constrained. In particular, for $m'_E\sim4$ TeV, the coupling to the third-generation SM lepton $\ell_L^\tau=(\nu_L^\tau,\tau_L)^T$ can be as large as $\mO(1)$ since the $Z\tau\tau$ and $W\tau\nu$ constraints are not very tight~\cite{ALEPH:2005ab,ParticleDataGroup:2024cfk}. Due to the mass hierarchy \Eq{mass_hierarchy} among the second generation VLLs, $E'$ can be first produced abundantly and then cascade decay to $N'$, whose decay to the SM particles triggers the leptogenesis. In the rest frame of the bubble wall, its vicinity can be approximated as a plane at $z=0$, with the $z$-axis pointing toward the true vacuum. Then the vertex $-\lambda_D^\tau\bar\ell_L^\tau\hat{h}E'_R\subset \mL'_{\nu\text{-VLL}}$ can be expanded as
\begin{align}
-\frac{\lambda_D^\tau}{\sqrt{2}}\bar\tau_L h(z)E'_R- \frac{\lambda_D^\tau}{\sqrt{2}}\bar\tau_L \hat hE'_R +\hc\,,
\end{align}
where $\hat{h}$ is the Higgs fluctuation field on top of the static background $h(z)$ with $h(\infty)=0$ and $h(-\infty)=v_{n1}$. $E'$ can be produced via two types of processes. The first one is the $1\to 2$ splitting $\tau_L\rightarrow E'_R \hat h$ or $\hat h\rightarrow E'_R \bar{\tau}_L$ via the $\hat h$ coupling, and we have checked the yield is negligible. The second is the $1\to 1$ mixing transition $\tau_L\rightarrow E'_R$ induced by the background field~\cite{Azatov:2020ufh}, which is dominant and will be discussed below.

The number density of the produced $E'$'s can be computed using the Bödeker-Moore method~\cite{Bodeker:2017cim}. Working in the wall frame, we can express the four-momenta as
\begin{subequations}
\label{eq:kinematics}
\begin{align}
    &{\rm incoming\
    }\tau_L:\quad\ \ p=(p^0,\vecp_\perp,p^z(z))\,,\\
    &{\rm outgoing\ }  E'_R: \quad\  k=(k^0,\veck_{\perp},k^z(z))\,,
\end{align}
\end{subequations}
where $\perp$ denotes the direction perpendicular to the $z$-axis, and on-shell conditions are understood. Although the $\tau$ and $E'$ masses are $z$-dependent due to interactions with $h(z)$, the hierarchy $m_\tau \ll v_{n1}\sim T_{n1} \ll m'_E$ allows us to approximate the $\tau_L$ as massless and the $E'_R$ mass as constant. Consequently, $p^z$ and $k^z$ are effectively $z$-independent, and the breaking of $z$-translation invariance originates entirely from the bubble wall profile in the vertex. Since particles are highly boosted to the $z^+$-direction, we take $p^z$, $k^z\geqslant 0$. 

In the wall frame, the $E'_R$ flux injected to the $z>0$ region reads 
\be
\label{eq:particle-number}
J_{E'_R}^{\rm wl}=\int\frac{\d^3 \vecp}{(2\pi)^3} \frac{p^z}{p^0} f^{\rm wl}_{\tau_L}(\vecp)\times \d\mathbb{P}_{\tau_L\rightarrow E'_R}(\vecp)\,,
\ee
where the $\tau_L$ outside the wall obeys the Lorentz-boosted Fermi-Dirac distribution, namely
\be
f_{\tau_L}^{\rm wl}(\vecp)=\frac{1}{e^{\gamma_w(|\p|-v_wp^z)/T_{n1}}+1}\,,
\ee
and $\d \mathbb{P}$ is the differential transformation probability
\be
\label{eq:dP}
\d \mathbb{P}_{\tau_L\rightarrow E'_R}(\vecp)=\frac{1}{2p^z}  \int\frac{\d^3 \veck}{(2\pi)^3\, 2k^0}\times(2\pi)^3 \delta(p^0-k^0)\delta^2(\vecp_\perp-\veck_\perp)|\mM|^2\,,
\ee
with $|\mM|^2$ being the squared invariant transition amplitude. In the plasma frame, the flux is $J_{E'_R}^{\rm pl}=J_{E'_R}^{\rm wl}/\gamma_w$, while the number density of the produced $E'_R$ is $n_{E'_R}^{\rm pl}=J_{E'_R}^{\rm pl}/v_w$. The integration over $\veck$ can be trivially carried out, and we reach
\begin{align}\label{eq:production-rate}
n_{E'_R}^{\rm pl}= &\frac{1}{v_w\gamma_w}\int\frac{\d^3 \vecp}{(2\pi)^3} \frac{1}{2 p^0} f^{\rm wl}_{\tau_L}(\vecp) \frac{|\mM|^2}{2\sqrt{(p^0)^2-\vecp_\perp^2 - (m'_E)^2}}\,.
\end{align} 
Given $\gamma_w\gg 1$, we have
\begin{align}
    \sqrt{|\vecp_\perp|^2+(p^z)^2}\approx p^z+\frac{1}{2} \frac{|\vecp_\perp|^2}{p^z}\,,\quad v_w\approx 1-\frac{1}{2\gamma_w^2}\,.
\end{align}
This means $f_{\tau_L}^{\rm wl}(\p)\approx \exp\left\{-\frac{1}{2T_{n1}}\left(\frac{\gamma_w\vecp_\perp^2 }{p^z}+\frac{p^z}{\gamma_w} \right) \right\}$ is strongly peaked at $p^z=\gamma_w |\vecp_\perp|$. Following Ref.~\cite{Ai:2023suz}, we can do the integral over $p^z$ using the method of steepest descent, and finally obtain
\begin{align}
\label{eq:nEL}
n_{E'_R}^{\rm pl}= \frac{\sqrt{2\pi T_{n1}}}{8\pi^2\gamma_w}\int_0^\infty\d |\vecp_\perp| e^{-|\vecp_\perp|/T_{n1}} \frac{|\vecp_\perp|^{1/2} |\mM|^2_{p^z=\gamma_w|\vecp_\perp|  } }{2\sqrt{(p^0)^2-\vecp_\perp^2 - (m'_E)^2}}  \,,
\end{align}
where the momenta now are 
\begin{subequations}
\begin{align}
&p =\Big(\sqrt{\gamma_w^2 +1}|\vecp_\perp|,\vecp_\perp,\gamma_w|\vecp_\perp|\Big)\,,\\
&k=\Big(\sqrt{\gamma_w^2+1}|\vecp_\perp|,-\vecp_\perp,\sqrt{\gamma_w^2|\vecp_\perp|^2  -(m'_E)^2 }\Big)\,.
\end{align}
\end{subequations}  

The last quantity that has not been determined yet is the transition amplitude of $\tau_L\to E'_R$, which can be written as~\cite{Bodeker:2017cim}
\begin{align}
\mM^{ss'}=-\sqrt{2} i \lambda_D^\tau p^0 \left(\frac{V_s}{A_s}-\frac{V_b}{A_b}\right) \bar{u}^s(k; E') P_R\,  v^{s'}(p; \tau)\,,
\end{align}
where $s$, $s'$ are spinor indices, $A_s= A_b= m'^2_E$, $V_s=0$, and $V_b=v_{n1}$. The amplitude squared gives
\begin{align}
|\mM|^2= \sum_{s,s'} \mM^{ss'} \left(\mM^{ss'}\right)^*=4 (\lambda_D^\tau)^2 p_0^2 \frac{v_{n1}^2}{m'^4_E} (k\cdot p)\approx 2(\lambda_D^\tau)^2\left(\frac{v_{n1}}{m'_E}\right)^2 \gamma_w^2 |\vecp_\perp|^2 \,.
\end{align}
Substituting the above into Eq.\,\eqref{eq:nEL}, we obtain
\be
n_{E'_R}^{\rm pl}=\frac{1}{2^{5/2}\pi^{3/2}}(\lambda_D^\tau)^2 \left(\frac{v_{n1}}{m'_E}\right)^2 T_{n1}^3\times \int_{m'_E/(\gamma_wT_{n1})}^\infty \d\xi\frac{\xi^{\frac{3}{2}} \, e^{-\xi}}{\sqrt{1-m'^2_E/(\gamma_wT_{n1}\xi)^2}}\,,
\ee
where $\xi=|\vecp_\perp|/T_{n1}$. The integral is of $\mO(1)$ and asymptotes to $3\sqrt{\pi}/4\approx 1.33$ for $\gamma_wT_{n1}\gg m'_E$, and we finally reach a neat result
\be\label{eq:nEL-final}
n_{E'_R}^{\rm pl}\approx \frac{3\sqrt{2}}{32\pi} (\lambda_D^\tau)^2 \left(\frac{v_{n1}}{m'_E}\right)^2 T_{n1}^3\approx 10^{-6}\times T_{n1}^3\left(\frac{\lambda_D^\tau}{0.1}\right)^2\left(\frac{v_{n1}}{T_{n1}}\right)^2 \left(\frac{20\,T_{n1}}{m'_E}\right)^2 \,.
\ee
We find $\gamma_w T_{n1}\gtrsim 20\,{\rm TeV}$ in the parameter space of interest, and the integral gives $1.34--1.36$ which is very close to the asymptotic value.

Right after production, $E'_R$ will quickly be redistributed evenly to $E'_R$ and $E'_L$ due to the Dirac mass term $m'_E$, resulting in the number densities of $n_{E'_R}^{\rm in}\approx n_{E'_L}^{\rm in}\approx n_{E'_R}^{\rm pl}/2$. Hereafter we use $n_{E'}^{\rm in}=n_{E'_L}^{\rm in}+n_{E'_R}^{\rm in}$ to represent the number density of the produced $E'$ in the plasma frame, and the superscript ``in'' means they are the initial value in the bubble interior for the second-stage evolution. Via the charge conjugation process $\bar\tau_L\to\bar E'$, an amount of $n_{\bar E'}^{\rm in}=n_{E'}^{\rm in}$ is also produced.

\subsection{Leptogenesis during the inverse EWPT}\label{subsec:leptogenesis}

The inverse EWPT efficiently produces $E'$ particles when the bubbles expand. Those $E'$'s are out-of-equilibrium and boosted by a factor of $\gamma_1=m'_E/T_{n1}\gg1$ in the plasma frame~\cite{Baldes:2021vyz}. They may decay via $E'\to F'H$, may annihilate with each other via $E'\bar E'\to HH^*$, $\ell^+\ell^-$, etc., or may thermalize with the plasma particles via elastic scattering with the EW-charged particles. Given $y'_{E_{L,R}}\sim1$, we have checked that the dominant process is decay,\footnote{When comparing the decay with thermalization, we have taken into account the time dilution factor $\gamma_1$ when transferring from $E'$-rest frame to the plasma frame.} therefore the $F'$ is produced with a number density of $n_{F'}^{\rm in}\approx n_{E'}^{\rm in}$, and it inherits the boost factor $\gamma_1$. This is not the end of the decay chain; as $y'_{N_{L,R}}\sim1$, $F'$ mainly decays via $F'\to N'H$ instead of annihilating or thermalizing, resulting in $N'$ with $n_{N'}^{\rm in}\approx n_{F'}^{\rm in}$ with the same boost factor $\gamma_1$. Being the lightest second-generation VLL, $N'$ can only decay or annihilate to SM particles via the seesaw vertices, and the thermalization rate is negligible since it is an EW singlet. Successful leptogenesis requires the fraction of decaying $N'$ is considerable, and the decay products exhibit a lepton asymmetry.

We then turn to the $N'$ decay process inside the bubble, where $h=0$ and the Higgs-portal Yukawa interactions do not induce any additional mass terms for the VLLs. In this case, the mass term for neutral fermions is
\be
\mL_{\rm mass}^{h\to 0}=-m'_N\bar N'_LN'_R-m'_F\bar F'^0_LF'^0_R-\frac{\mu_N}{2}\bar N'_LN'^c_L-\sum_\alpha\mu_D^\alpha\bar\nu_L^\alpha F'^0_R+\hc\,,
\ee
where the $\mu_D$-term induces direct $\ell_L$-$F'^0_R$ mixing, while the $\mu_N$-term splits $N'$ from a Dirac fermion into two near-degenerate Majorana fermions $\chi_{1,2}$. We first use the $4\times 4$ unitary transformation $U_F$ to eliminate the $\mu_D$-term via $(\ell_L^\alpha,\, F'_L)^T\to U_F(\ell_L^\alpha,\, F'_L)^T$, and then diagonalize the $N'$-mass matrix via
\be\label{UN2}
\begin{pmatrix}N'^c_L\\ N'_R\end{pmatrix}= U_N\begin{pmatrix}\chi_1\\ \chi_2\end{pmatrix},\quad U_N\approx\begin{pmatrix}\frac{i}{\sqrt{2}}\left(1-\frac{\mu_N}{4m'_N}\right) & \frac{1}{\sqrt{2}}\left(1+\frac{\mu_N}{4m'_N}\right) \\ -\frac{i}{\sqrt{2}}\left(1+\frac{\mu_N}{4m'_N}\right) & \frac{1}{\sqrt{2}}\left(1-\frac{\mu_N}{4m'_N}\right) \end{pmatrix},
\ee
such that
\be
\mL_{\rm mass}^{h\to 0}\supset-\frac12\begin{pmatrix}\bar N'_L & \bar N'^c_R\end{pmatrix}\begin{pmatrix}\mu_N & m'_N \\ m'_N & 0\end{pmatrix}\begin{pmatrix}N'^c_L\\ N'_R\end{pmatrix}+\hc\approx-\frac12\sum_{i=1}^2m_{\chi_i}\bar\chi_i\chi_i+\hc\,,
\ee
with $m_{\chi_{1,2}}\equiv m'_N\mp\mu_N/2$. Note that in the EW-breaking vacuum there are two pairs of near-degenerate Majorana heavy neutrinos, coming from $N'$ and $F'^0$ respectively, while here in the EW-symmetric vacuum there is only one pair of $\chi$. This is significantly different from the traditional inverse seesaw scenario, and the underlying reason is that $F'^0$ is not an EW singlet, such that terms like $\bar F'^{0c}_LF'^0_L$ cannot be written.

The two heavy neutrinos $\chi_{1,2}$ are the RHNs needed for leptogenesis, and their couplings with the SM left-handed leptons are given by $-\sum_{\alpha,i}h_\nu^{\alpha i}\bar\ell_L^\alpha\tilde H\chi_i$, where again $\alpha=e$, $\mu$, $\tau$, and $i=1$, 2, and explicitly
\be
h_\nu^{\alpha i}=\lambda_N(U_F^{4\alpha})^*U_N^{1i}+\left[y'_{N_R}(U_F^{4\alpha})^*+\sum_\beta y_D^\beta (U_F^{\beta\alpha})^*\right]U_N^{2i}\,.
\ee
The Yukawa matrix $h_\nu^{\alpha i}$ induces $\chi_i\to\ell^\alpha H/\bar\ell^\alpha H^*$ decays, which are the sources of leptogenesis. The flavored decay width asymmetry is defined as
\be
\epsilon_{\alpha i}=\frac{\Gamma_{\chi_i\to\ell^\alpha H}-\Gamma_{\chi_i\to\bar\ell^\alpha H^*}}{\sum_\beta\left(\Gamma_{\chi_i\to\ell^\beta H}+\Gamma_{\chi_i\to\bar\ell^\beta H^*}\right)}\,,
\ee
and it is zero at tree level, since $\Gamma^{\rm tree}_{\chi_i\to\ell^\alpha H}=\Gamma^{\rm tree}_{\chi_i\to\bar\ell^\alpha H^*}=|h_\nu^{\alpha i}|^2m_{\chi_i}/(16\pi)$. Nonzero $\epsilon_{\alpha i}$ can be induced by the CP phases in the $h_\nu$ couplings via the interference between tree and one-loop diagrams~\cite{Pilaftsis:1997jf,Pilaftsis:2003gt,Adhikary:2014qba}
\begin{multline}
\epsilon_{\alpha i}=\frac{1}{8\pi(h_\nu^\dagger h_\nu)_{ii}}\sum_{j\neq i}\text{Im}[(h_\nu^\dagger h_\nu)_{ij}(h_\nu^\dagger)_{i\alpha}h_\nu^{\alpha j}]\left[f(x_{ij})+\frac{\sqrt{x_{ij}}(1-x_{ij})}{(1-x_{ij})^2+\frac{1}{64\pi^2}(h_\nu^\dagger h_\nu)_{jj}^2}\right]\\
+\frac{1}{8\pi(h_\nu^\dagger h_\nu)_{ii}}\sum_{j\neq i}\frac{(1-x_{ij})\text{Im}[(h_\nu^\dagger h)_{ji}(h_\nu^\dagger)_{i\alpha}h_\nu^{\alpha j}]}{(1-x_{ij})^2+\frac{1}{64\pi^2}(h_\nu^\dagger h_\nu)_{jj}^2}+\mO(h_\nu^6)\,,
\end{multline}
with $x_{ij}=(m_{\chi_j}/m_{\chi_i})^2$ and $f(x)=\sqrt{x}\,[1-(1+x)\ln(1+1/x)]$. We further denote the unflavored decay asymmetry as $\epsilon_i=\sum_\alpha\epsilon_{\alpha i}$.

The CP-violating effects are induced by the low-energy phases $\delta_{\rm CP}$ and $\alpha_{21}$ in the PMNS matrix, and the high-energy phase in the complex $y_D$ and $\mu_D$ matrices. Current neutrino oscillation data imply a large $\delta_{\rm CP}\sim3\pi/2$~\cite{ParticleDataGroup:2024cfk}, while the Majorana and high-energy phases have yet to be measured experimentally. However, the low-energy phases cancel out in the unflavored decay asymmetries, and it is challenging to realize flavored leptogenesis solely via $\delta_{\rm CP}$~\cite{Branco:2011zb}.\footnote{For recent trials in the thermal leptogenesis framework using pseudo-Dirac RHNs, see for example Refs.~\cite{Dolan:2018qpy,Li:2021tlv,Mukherjee:2023nyi,Shao:2025lym}.} Our scenario belongs to unflavored leptogenesis, and the nonzero $\epsilon_i$ is sourced by the high-energy phase. In the CI parametrization scheme in \Eq{CI}, the phase comes from the complex numbers in $\mu^{-1/2}$, requiring nonzero $\mu_N$ and $\lambda_N$~\cite{Deppisch:2010fr}.

With $\epsilon_i$ in hand, we now investigate the fate of $\chi_i$'s after being produced from $F'$ decay. The initial number densities of $\chi_{1,2}$ inherit from the bubble-produced $E'$, yielding $n_{\chi_1}^{\rm in}\approx n_{\chi_2}^{\rm in}\approx n_{E'}^{\rm in}$. The $\chi$'s are boosted by a Lorentz factor of $\gamma_1\sim m'_E/T_{n1}$ in the plasma frame, but they are almost at rest in the $\chi$-gas frame, in which the number densities are $n_{\chi}^{\rm gas}\approx\gamma_1n_{E'}^{\rm in}$. In this frame, $\chi$-fermions can decay or annihilate, and we can track the evolution via the following simplified equation:
\be
\dot{\mathfrak{n}}_\chi(t)=-\Gamma_\chi \mathfrak{n}_\chi(t)-\ave{\sigma v}\mathfrak{n}_\chi^2(t),\quad \mathfrak{n}_\chi(0)=n_\chi^{\rm gas},
\ee
where the cosmic expansion term is tiny and hence dropped. The solution is
\be
\mathfrak{n}_\chi(t)=\frac{n_\chi^{\rm gas}e^{-\Gamma_\chi t}}{1+(\Gamma_{\rm ann}/\Gamma_\chi)(1-e^{-\Gamma_\chi t})}\,,
\ee
where $\Gamma_{\rm ann}=n_\chi^{\rm gas}\ave{\sigma v}$ is the annihilation rate. During this evolution, the fraction of particles that experience decay is
\be
f_{\rm dec}=\frac{1}{n_\chi^{\rm gas}}\int_0^\infty\d t'\mathfrak{n}_\chi(t)\Gamma_\chi=\frac{\Gamma_\chi}{\Gamma_{\rm ann}}\log\left(1+\frac{\Gamma_{\rm ann}}{\Gamma_\chi}\right).
\ee
It is easy to check $f_{\rm dec}\to0$ for $\Gamma_{\rm ann}\gg\Gamma_\chi$ and $f_{\rm dec}\to 1$ for $\Gamma_{\rm ann}\ll\Gamma_\chi$.

Apparently, we need a sizable $f_{\rm dec}$ for leptogenesis. In our scenario, $\Gamma_\chi\approx \Gamma^{\rm tree}_{\chi_i\to\ell^\alpha H}+\Gamma^{\rm tree}_{\chi_i\to\bar\ell^\alpha H^*}$, while $\Gamma_{\rm ann}$ is dominated by $\ave{\sigma_{\chi_1\chi_2\to HH^*} v}\approx m'^2_N(y'^2_{N_L}+y'^2_{N_R})^2/[32\pi(m'^2_N+m'^2_F)^2]$. Although the $2\to2$ annihilation suffers from the phase space suppression compared with the $1\to2$ decay, $\Gamma_{\rm ann}$ might be comparable or even larger than $\Gamma_\chi$ due to the $y'_{N_{L,R}}\sim\mO(1)$ couplings responsible for annihilation and the small seesaw vertices $|h_\nu^{\alpha i}|\sim\mO(10^{-5})$ responsible for decay. We emphasize that the ``decay versus annihilation'' issue is general in all low-scale bubble-assisted leptogenesis mechanisms, as the decay vertices come from the seesaw couplings, which are quite small at TeV scale.\footnote{Ref.~\cite{Dasgupta:2022isg} studies bubble-assisted leptogenesis based on the mass-gain mechanism during a $U(1)_{B-L}$ FOPT at TeV scale within the type-I seesaw framework. However, the authors ignore the annihilation processes, leading to a significantly overestimated BAU.} In fact, to enhance $f_{\rm dec}$, we adopt the inverse seesaw instead of the more studied type-I seesaw, as the corresponding vertices in the former are enhanced compared with the latter.

Since the $\chi_i$'s are boosted, the decay products of $\chi_i$ are also boosted. We should make sure the $\ell/\bar\ell$ and $H/H^*$ particles thermalize with the plasma instead of fusion back to $\chi_i$, which will wash the $B-L$ number away. This means $\Gamma_{\rm th}>H(T_{n1})$ and $\Gamma_{\rm th}>\Gamma_{\rm id}$, where~\cite{Huang:2022vkf}
\be\label{thermalization}
\Gamma_{\rm th}=\frac{2\zeta_3g_{\rm EW}\alpha_W^2T_{n1}^3}{4\pi m'^2_N}\log\frac{3m'^2_N}{5\pi\cdot 2\alpha_WT_{n1}^2}\,,\quad \Gamma_{\rm id}=\frac{2^2T_{n1}^3}{4m'^3_N}\left(\sum_i\Gamma_{\chi_i}\right)e^{-2^2/4}\,,
\ee
where $g_{\rm EW}=46$ is the number of d.o.f. of the EW-charged SM fermions. Once the above inequalities are satisfied, the $B-L$ asymmetry generated by $\chi_i$ decay will quickly thermalize, and we are faced with the final suppression factor, which is the thermal washout from the inverse decay $\ell H\to\chi_i\to\ell H$ and its charge conjugation in the plasma, governed by the Boltzmann equation
\be
\frac{\d Y_{B-L}}{\d z}+\frac{z^4}{s_NH_N}\frac{Y_{B-L}}{2Y_\ell^{\rm eq}}\sum_i\gamma_{\chi_i}=0\,,
\ee
where $z=m'_N/T$, and the decay rate is defined as
\be
\gamma_{\chi_i}\approx(\Gamma_{\chi_i\to\ell H}+\Gamma_{\chi_i\to\bar\ell H^*})\frac{m_{\chi_i}^3}{\pi^2}\frac{K_1(z)}{z}\equiv\gamma_{\chi_i\to\ell H}+\gamma_{\chi_i\to\bar\ell H^*}
\ee
with $K_1$ being the modified Bessel function of the first kind, and the partial widths are $\gamma_{\chi_i\to\ell H}=(1+\epsilon_i)\gamma_{\chi_i}/2$, and $\gamma_{\chi_i\to\bar\ell H^*}=(1-\epsilon_i)\gamma_{\chi_i}/2$.

Combining all physical effects above, i.e., the cascade decay of $E'$, the decay fraction and decay asymmetry of $\chi$, and the thermal washout, the final BAU generated from the bubble-assisted leptogenesis is
\be\label{YB_bubble}
Y_B^{\rm bub.}=-\frac{28}{79}\sum_{i=1}^2\epsilon_i\frac{f_{\rm dec}n_\chi^{\rm in}}{2\pi^2g_*T_{n1}^3/45}\exp\left\{-\int_{z_{\rm in}}^{z_{\rm di}}\frac{z'^4\d z'}{s_NH_N}\frac{\sum_{j=1}^2\gamma_{\chi_j}(z')}{2Y_\ell^{\rm eq}}\right\},
\ee
where the prefactor comes from the conversion efficiency of the EW sphaleron, and the $z$-evolution starts from $z_{\rm in}=m'_N/T_{n1}$ (the inverse EWPT), while ends at $z_{\rm di}=m'_N/T_{n2}$ (the later direct EWPT, which terminates the EW sphaleron). The variables in the denominator of the exponent are $s_N=2\pi^2g_*m'^3_N/45$, $H_N=2\pi m'^2_N\sqrt{\pi g_*/45}/M_{\rm Pl}$, and $Y_\ell^{\rm eq}=135\zeta_3/(4\pi^2g_*)$, where $M_{\rm Pl}=1.22\times10^{19}$ GeV is the Planck scale. Given $m'_N\ll M_{\rm Pl}$, the washout factor is typically large, rendering $Y_B$ negligible, unless the $K_1(m'_N/T)\sim e^{-m'_N/T}$ term from $\gamma_\chi$ is sufficiently suppressed. Numerically, this suppression requires $m'_N/T_{n_1}\gtrsim20$, a number mentioned in the Introduction. We emphasize that this issue is common in low-scale bubble-assisted leptogenesis; in this work, we introduce a second generation of VLLs with masses at $\sim4~{\rm TeV}$ to achieve this condition. Unlike the previous bubble-assisted leptogenesis studies, our BAU in \Eq{YB_bubble} does not suffer from the FOPT reheat suppression, because the EWPTs here are not supercooled.

\begin{figure}
    \centering
    \includegraphics[scale=0.55]{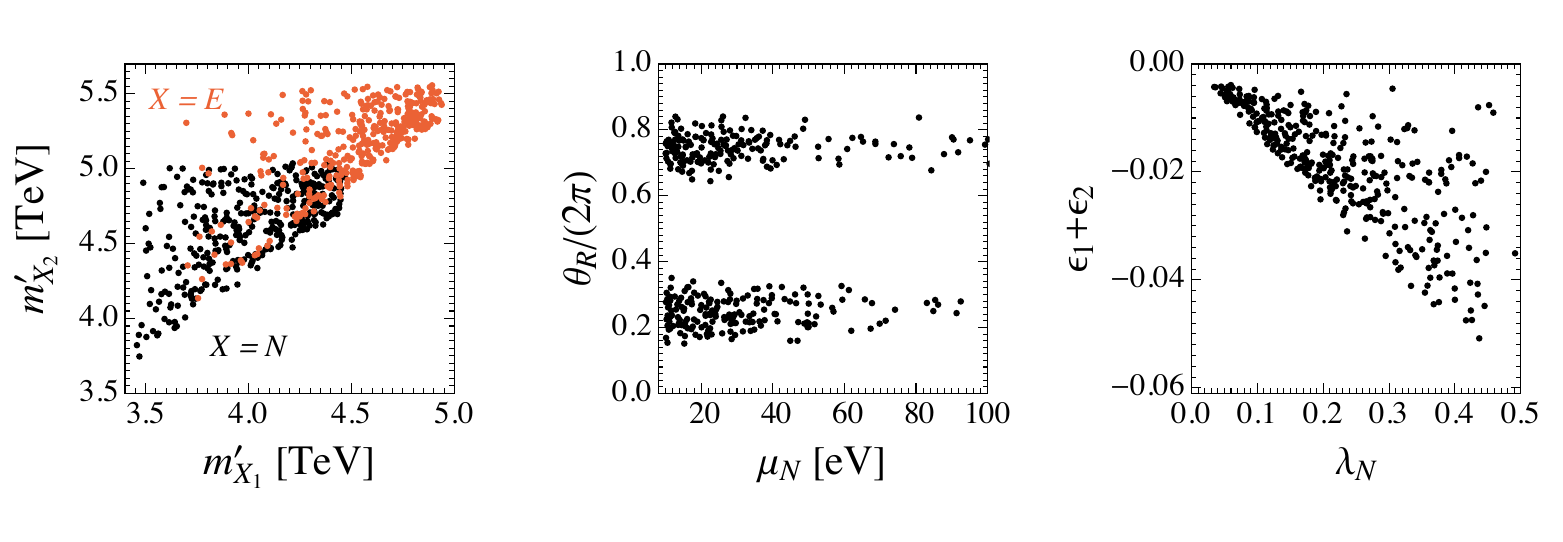}
    \caption{The bubble-assisted leptogenesis parameter space projected to two-dimensional plots. Left: the physical masses $m'_{X_1}$ v.s. $m'_{X_2}$ with $X=N$ and $E$. Middle: the Majorana mass term $\mu_N$ v.s. the $\theta_R$ angle in the CI parametrization. Right: the lepton-number-breaking Yukawa coupling $\lambda_N$ v.s. the decay asymmetries $\epsilon_1+\epsilon_2$ in the EW-symmetric vacuum.}
    \label{fig:leptogenesis_parameter_space}
\end{figure}

Based on the inverse EWPT parameter space obtained in section~\ref{subsec:ewpt}, we scan over
\be\begin{split}
&m'_N\in[3.5~{\rm TeV},\,4.5~{\rm TeV}]\,,\quad m'_F\in[m'_N,\,5~{\rm TeV}]\,,\quad m'_E\in[m'_F,\,5.5~{\rm TeV}]\,;\\
&y'_{N_{L,R}}, \, y'_{E_{L,R}}\in[0.5,\,1.5]\,;\\
&\theta_R\in[0,\,2\pi]\,,\quad \log_{10}\lambda_N\in[-9,\,0]\,,\quad \log_{10}(\mu_N/{\rm GeV})\in[-8,\,2]\,,
\end{split}\ee
to find the parameter space for successful leptogenesis. During the scan, we adopt $\lambda_D^\tau\sim\mO(0.1)$, and its value is fixed by the requirement of $Y_B^{\rm bub.}=Y_B$. A larger $\lambda_D^\tau$ results in a larger production rate of $E'$, which can enhance $n_\chi^{\rm in}$; however, if $n_\chi^{\rm in}$ is too large, the annihilation rate also increases, reducing $f_{\rm dec}$ and hence $Y_B^{\rm bub.}$. The results are shown in Fig.~\ref{fig:leptogenesis_parameter_space}, and we verify that the conventional thermal leptogenesis suffers from large thermal washout effects, yielding $Y_B^{\rm th}\lesssim10^{-13}$. Therefore, this is the parameter space where our nonthermal leptogenesis mechanism dominates and successfully accounts for the BAU.

The left panel of Fig.~\ref{fig:leptogenesis_parameter_space} displays the mass eigenstates $m'_{N_{1,2}}$ and $m'_{E_{1,2}}$ of the second-generation VLLs. The Higgs portal Yukawa couplings $y'_{N_{L,R}}$ and $y'_{E_{L,R}}$ distribute nearly uniformly in the scanned region, and hence are not shown. Instead, we present the distribution of the Majorana mass term of $N'_L$ and the $\theta_R$ angle in the CI parametrization in the middle panel, which shows that $\mu_N$ favors the $\mO(10~{\rm eV})$ range and $\theta_R$ is mainly around $\sim 0.2\times 2\pi$ and $\sim0.8\times 2\pi$. We also obtain that the unflavored decay asymmetries satisfy $\epsilon_1\approx\epsilon_2$, and they are most sensitive to the lepton-number-violating Yukawa coupling $\lambda_N$ between $F'_L$ and $N'^c_L$. The right panel then plots the $\sum_i\epsilon_i$ and $\lambda_N$, revealing a correlation. Although the scan covers $\lambda_N \geqslant 10^{-9}$, the requirement from leptogenesis constrains $\lambda_N\gtrsim0.01$. Besides, we find that $|y_D^\alpha|\sim\mO(10^{-5})$ and $|\mu_D^\alpha|\sim\mO({\rm MeV})$.

\section{Conclusion}\label{sec:conclusion}

In this work, we propose a novel nonthermal leptogenesis mechanism by extending the SM with two generations of VLLs. The first-generation VLLs help the Higgs field to realize an inverse EWPT during the cosmic cooling, causing relativistic bubble expansion. The second-generation VLLs are embedded into the minimal inverse seesaw mechanism to explain the SM neutrino mass origin. When the bubbles expand, SM light particles hit the fast-moving bubble walls and abundantly produce the second-generation VLLs, which ultimately decay to the SM leptons to realize leptogenesis.

To the best of our knowledge, our work presents the first bubble-assisted leptogenesis mechanism at an EWPT. Achieving this is highly nontrivial. First, to ensure active EW sphalerons, we consider an unconventional inverse EWPT. Second, for TeV-scale RHNs, the smallness of the seesaw couplings means that annihilation can dominate over decay, and hence no lepton asymmetry is generated. To circumvent this, we employ the inverse seesaw instead of the type-I seesaw, which enhances the decay rate. Third, the decay products of the RHN rapidly thermalize and are subject to significant thermal washout effects at EW-scale temperatures. We suppress this by setting the second-generation VLL mass to be around 4 TeV. We emphasize here that the three challenges outlined above, namely maintaining active sphalerons, ensuring decay dominates over annihilation, and suppressing thermal washout, are common for all TeV-scale bubble-assisted leptogenesis mechanisms. Notably, the latter two effects have been overlooked in certain previous studies, as discussed in the main text.

\begin{figure}
    \centering
    \includegraphics[scale=0.55]{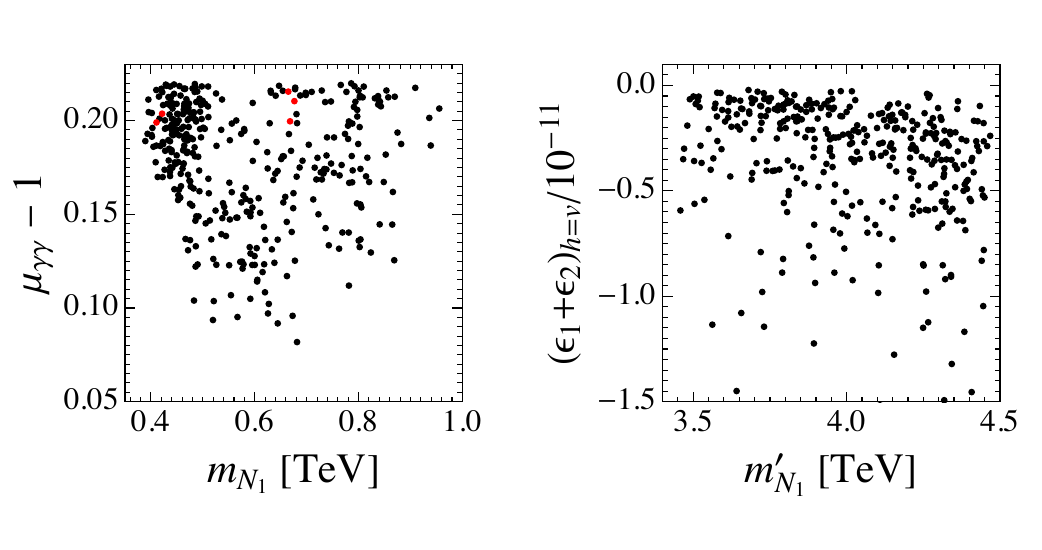}
    \caption{Left: the physical mass $m_{N_1}$ v.s. the $h\to\gamma\gamma$ signal strength deviation. The red points yield a SNR $>1$ at the GW detector BBO. Right: the physical mass $m'_{N_1}$ v.s. the decay asymmetries in the EW-breaking vacuum.}
    \label{fig:phenomenology}
\end{figure}

Our model can be tested by current and future experiments. The first-generation VLLs are around 1 TeV, and hence can be probed directly at the LHC~\cite{ATLAS:2024mrr,ATLAS:2025wgc}. Furthermore, the charged VLLs modify the Higgs diphoton decay rate, and the conditions for an inverse EWPT typically require a signal strength of $\mu_{\gamma\gamma} > 1.05$, as shown in the left panel of Fig.~\ref{fig:phenomenology}. Such a significant deviation is well within the reach of the future high-luminosity LHC~\cite{Liss:2013hbb,CMS:2013xfa,ATL-PHYS-PUB-2014-016}. Another potential signature is the stochastic GWs from the EWPTs. However, as indicated in Fig.~\ref{fig:EWPT_parameters}, both EWPTs are characterized by a weak strength ($\alpha \lesssim 10^{-2}$) and a rapid transition rate ($\beta/H_n \gtrsim 10^4$). Consequently, the resultant GW signal is weak, and only a small fraction of the parameter space yields a signal-to-noise ratio (SNR) greater than 1 at the future Big Bang Observer (BBO) detector~\cite{Crowder:2005nr}, assuming a 10-year data-taking period, as marked by red points in the plot.

Although the second-generation VLLs are around four times heavier than the first-generation ones, they are still available at the future 10 TeV muon collider~\cite{Guo:2023jkz}. The large CP asymmetries $|\epsilon_i|\sim\mO(10^{-2})$ shown in the right panel of Fig.~\ref{fig:leptogenesis_parameter_space} might suggest they could be measured via charged lepton asymmetries from RHN decays at the future muon collider~\cite{Liu:2021akf}. However, a crucial distinction must be made: the $\epsilon_i$ relevant for leptogenesis is computed in the EW-symmetric vacuum ($h=0$), whereas particle experiments probe the asymmetries in the EW-breaking vacuum ($h=v$). As demonstrated in the right panel of Fig.~\ref{fig:phenomenology}, the latter are too small to be detected. This suppression arises because the CP asymmetries are sensitive to the mass splitting between the RHNs, which differs significantly between the two vacuum states. Despite this specific challenge, the model overall predicts a wide range of observable phenomena, and we leave the detailed studies for future work.

\acknowledgments

We would like to thank Giulio Barni, Tomasz Dutka, Claudia Hagedorn, Shao-Ping Li, Yan Shao, Miguel Vanvlasselaer, Shao-Jiang Wang, Zhen-hua Zhao, and Ye-Ling Zhou for the helpful discussions and comments. The work of WA  is supported by the European Union (ERC, NLO-DM, 101044443). The work of PH is supported by the National Science Foundation under grant number PHY-2412875, and the University Research Association Visiting Scholars Program. PH is grateful for the hospitality of Perimeter Institute where part of this work was carried out. Research
at Perimeter Institute is supported in part by the Government of Canada through the Department of
Innovation, Science and Economic Development and by the Province of Ontario through the Ministry of
Colleges and Universities. This work was supported by a grant from the Simons Foundation (1034867,
Dittrich). KPX is supported by the National Science Foundation of China under Grant No. 12305108.

\appendix
\section{Finite-temperature dynamics and EWPTs}\label{app:ewpt}

The fully one-loop effective Higgs potential $V_T(h)$ consists of the zero-temperature potential and thermal corrections. To evaluate these, we first derive the field-dependent mass squared of the particles in our model. For the VLLs, $\mM_{N_{1,2}}^2(h)$ and $\mM_{E_{1,2}}^2(h)$ are obtained by the replacement $v\to h$ in \Eq{MX12}. For the SM particles, we get
\be\begin{split}
\mM_W^2(h)=&~\frac{m_W^2}{v^2}h^2\,,\quad \mM_Z^2(h)=\frac{m_Z^2}{v^2}h^2\,,\quad \mM_t^2(h)=\frac{m_t^2}{v^2}h^2\,,\\
\mM_h^2(h)=&~\frac{m_h^2}{2v^2}(3h^2-v^2)\,,\quad \mM_{G^{\pm,0}}^2(h)=\frac{m_h^2}{v^2}(h^2-v^2)\,, 
\end{split}\ee
for the $W^\pm/Z$ bosons, the top quark, the Higgs boson, and the Goldstone bosons $G^{\pm,0}$, with $m_W=80.4$ GeV, $m_Z=91.2$ GeV, and $m_t=173$ GeV being the physical masses.

The one-loop Coleman-Weinberg potential~\cite{Coleman:1973jx} is
\be
V_{\rm CW}(h)=\sum_{i=W,Z,h,G,t,N_{1,2},E_{1,2}}\frac{n_i\mM_i^4(h)}{64\pi^2}\left(\log\frac{\mM_i^2(h)}{m_i^2}-\frac32\right),
\ee
with $n_{W}=6$, $n_Z=3$, $n_h=1$, $n_G=3$, $n_t=-12$, $n_{N_{1,2}}=n_{E_{1,2}}=-4$. The counter term
\be
\delta V(h)=\frac{\delta\mu^2}{2}h^2+\frac{\delta\lambda}{4}h^4
\ee
is introduced to make
\be
\frac{\d}{\d h}(V_{\rm CW}+\delta V)\Big|_{h=v}=0\,,\quad \frac{\d^2}{\d h^2}(V_{\rm CW}+\delta V)\Big|_{h=v}=0\,,
\ee
such that the one-loop zero-temperature potential
\be
V_1(h)\equiv V_0(h)+V_{\rm CW}(h)+\delta V(h)
\ee
still has the VEV at $v$ with the Higgs mass boson mass $m_h$.

In the early Universe, when the space is filled by a thermal bath with temperature $T$, the one-loop thermal corrections read
\be
V_{1,T}(h,T)=\sum_{i=W,Z,h,G}\frac{n_iT^4}{2\pi^2}J_B\left(\frac{M_i^2(h^2)}{T^2}\right)+\sum_{i=t,N_{1,2},E_{1,2}}\frac{n_iT^4}{2\pi^2}J_F\left(\frac{M_i^2(h^2)}{T^2}\right),
\ee
where the thermal integrals are
\be
J_{B,F}(y)=\int_0^\infty x^2\d x\log\left(1\mp e^{-\sqrt{x^2+y}}\right).
\ee
In addition, we include the daisy resummation to fix the IR divergence at small couplings,
\be
V_{\rm daisy}(h,T)=-\sum_{i=W,Z,\gamma,h,G}\frac{\bar n_iT}{12\pi}\left[(M_i^2(h)+\Pi_i(T))^{3/2}-M_i^3(h)\right],
\ee
where $\bar n_{W}=2$, $\bar n_Z=1$, $\bar n_\gamma=1$, $\bar n_h=1$, and $\bar n_{G}=3$, and the Debye masses are
\be\begin{split}
\Pi_{h,G}=&~\frac{T^2}{4v^2}(m_h^2+m_Z^2+2m_W^2+2m_t^2)\,,\quad \Pi_W(T)=\frac{22T^2}{3v^2}m_W^2\,,\\
\Pi_{Z}(T)=&~\frac{22T^2}{3v^2}(m_Z^2-m_W^2)-\mM_W^2(h)\,,\quad \Pi_\gamma(T)=\mM_W^2(h)+\frac{22T^2}{3v^2}m_W^2\,.
\end{split}\ee
Combining the information above, we reach the fully one-loop thermal potential
\be\label{VT}
V_T(h,T)=V_1(h)+V_{1,T}(h,T)+V_{\rm daisy}(h,T)\,.
\ee

As first realized in Ref.~\cite{Angelescu:2018dkk}, the potential in \Eq{VT} may experience a special thermal history. When $T$ drops during the cosmic expansion, the Higgs VEV first shifts to a nonzero value via a smooth crossover, then tunnels back to the field origin via an inverse FOPT, and then tunnels to the nonzero EW-breaking vacuum again via a direct FOPT, leading to a double-EWPT scenario. During either EWPT period, the potential has two degenerate local minima (vacua) at the critical temperature $T_c$, while below $T_c$ the true vacuum (global minimum) has a lower free energy. The decay rate from the false vacuum to the true vacuum per unit volume is
\be
\Gamma(T)\approx T^4\left(\frac{S_3}{2\pi T}\right)^{3/2}e^{-S_3/T},
\ee
where $S_3$ is the Euclidean action of the $O(3)$-symmetric bounce solution~\cite{Linde:1981zj}. The nucleation temperature $T_n$ is resolved from $\Gamma(T_n)H^{-4}(T_n)=1$, which means the probability of vacuum decay in a Hubble patch during a Hubble time reaches $\mO(1)$, where the Hubble constant is $H(T)=2\pi\sqrt{\pi g_*/45}(T^2/M_{\rm Pl})$.  Numerically, this yields $S_3/T_n\sim140$ for an EWPT~\cite{Quiros:1999jp}. The false vacuum volume fraction is given by
\be
p(T)=\exp\left\{-\frac{4\pi}{3}\int_T^{T_c}\frac{\d T'}{T'^4}\frac{\Gamma(T')}{H(T')}\left(\int_T^{T'}\d T''\frac{v_w}{H(T'')}\right)^3\right\},
\ee
and the percolation temperature satisfies $p(T_{p})=0.71$, where the true vacuum bubbles form an infinite connected cluster in the cosmic space. We use the numbers 1 and 2 to label the inverse and direct EWPTs, respectively, and the characteristic temperatures are denoted as $T_{c1,c2}$, $T_{n1,n2}$, $T_{p1,p2}$, etc. Since the EWPTs in this model are not supercooled, percolation is very close to nucleation, and we use $T_n$ as the transition temperature.

\bibliographystyle{JHEP-2-2.bst}
\bibliography{references}

\providecommand{\href}[2]{#2}\begingroup\raggedright\begin{thebibliography}{10}

\bibitem{ParticleDataGroup:2024cfk}
{\scshape Particle Data Group} collaboration, S.~Navas et~al., ``{Review of
  particle
  physics},''\href{http://dx.doi.org/10.1103/PhysRevD.110.030001}{\emph{Phys.
  Rev. D} {\bf 110} (2024) 030001}.

\bibitem{Kuzmin:1985mm}
V.~A. Kuzmin, V.~A. Rubakov and M.~E. Shaposhnikov, ``{On the Anomalous
  Electroweak Baryon Number Nonconservation in the Early
  Universe},''\href{http://dx.doi.org/10.1016/0370-2693(85)91028-7}{\emph{Phys.
  Lett. B} {\bf 155} (1985) 36}.

\bibitem{Minkowski:1977sc}
P.~Minkowski, ``{$\mu \to e\gamma$ at a Rate of One Out of $10^{9}$ Muon
  Decays?},''\href{http://dx.doi.org/10.1016/0370-2693(77)90435-X}{\emph{Phys.
  Lett. B} {\bf 67} (1977) 421--428}.

\bibitem{Buchmuller:2004nz}
W.~Buchmuller, P.~Di~Bari and M.~Plumacher, ``{Leptogenesis for
  pedestrians},''\href{http://dx.doi.org/10.1016/j.aop.2004.02.003}{\emph{Annals
  Phys.} {\bf 315} (2005) 305--351},
  [\href{https://arxiv.org/abs/hep-ph/0401240}{{\tt hep-ph/0401240}}].

\bibitem{Buchmuller:2005eh}
W.~Buchmuller, R.~D. Peccei and T.~Yanagida, ``{Leptogenesis as the origin of
  matter},''\href{http://dx.doi.org/10.1146/annurev.nucl.55.090704.151558}{\emph{Ann.
  Rev. Nucl. Part. Sci.} {\bf 55} (2005) 311--355},
  [\href{https://arxiv.org/abs/hep-ph/0502169}{{\tt hep-ph/0502169}}].

\bibitem{Davidson:2008bu}
S.~Davidson, E.~Nardi and Y.~Nir,
  ``{Leptogenesis},''\href{http://dx.doi.org/10.1016/j.physrep.2008.06.002}{\emph{Phys.
  Rept.} {\bf 466} (2008) 105--177},
  [\href{https://arxiv.org/abs/0802.2962}{{\tt 0802.2962}}].

\bibitem{Pilaftsis:2009pk}
A.~Pilaftsis, ``{The Little Review on
  Leptogenesis},''\href{http://dx.doi.org/10.1088/1742-6596/171/1/012017}{\emph{J.
  Phys. Conf. Ser.} {\bf 171} (2009) 012017},
  [\href{https://arxiv.org/abs/0904.1182}{{\tt 0904.1182}}].

\bibitem{Xing:2020ald}
Z.-z. Xing and Z.-h. Zhao, ``{The minimal seesaw and leptogenesis
  models},''\href{http://dx.doi.org/10.1088/1361-6633/abf086}{\emph{Rept. Prog.
  Phys.} {\bf 84} (2021) 066201}, [\href{https://arxiv.org/abs/2008.12090}{{\tt
  2008.12090}}].

\bibitem{Fukugita:1986hr}
M.~Fukugita and T.~Yanagida, ``{Baryogenesis Without Grand
  Unification},''\href{http://dx.doi.org/10.1016/0370-2693(86)91126-3}{\emph{Phys.
  Lett. B} {\bf 174} (1986) 45--47}.

\bibitem{Luty:1992un}
M.~A. Luty, ``{Baryogenesis via
  leptogenesis},''\href{http://dx.doi.org/10.1103/PhysRevD.45.455}{\emph{Phys.
  Rev. D} {\bf 45} (1992) 455--465}.

\bibitem{Lazarides:1990huy}
G.~Lazarides and Q.~Shafi, ``{Origin of matter in the inflationary
  cosmology},''\href{http://dx.doi.org/10.1016/0370-2693(91)91090-I}{\emph{Phys.
  Lett. B} {\bf 258} (1991) 305--309}.

\bibitem{Asaka:1999jb}
T.~Asaka, K.~Hamaguchi, M.~Kawasaki and T.~Yanagida, ``{Leptogenesis in
  inflationary
  universe},''\href{http://dx.doi.org/10.1103/PhysRevD.61.083512}{\emph{Phys.
  Rev. D} {\bf 61} (2000) 083512},
  [\href{https://arxiv.org/abs/hep-ph/9907559}{{\tt hep-ph/9907559}}].

\bibitem{Giudice:1999fb}
G.~F. Giudice, M.~Peloso, A.~Riotto and I.~Tkachev, ``{Production of massive
  fermions at preheating and
  leptogenesis},''\href{http://dx.doi.org/10.1088/1126-6708/1999/08/014}{\emph{JHEP}
  {\bf 08} (1999) 014}, [\href{https://arxiv.org/abs/hep-ph/9905242}{{\tt
  hep-ph/9905242}}].

\bibitem{Mazumdar:2018dfl}
A.~Mazumdar and G.~White, ``{Review of cosmic phase transitions: their
  significance and experimental
  signatures},''\href{http://dx.doi.org/10.1088/1361-6633/ab1f55}{\emph{Rept.
  Prog. Phys.} {\bf 82} (2019) 076901},
  [\href{https://arxiv.org/abs/1811.01948}{{\tt 1811.01948}}].

\bibitem{Caprini:2019egz}
C.~Caprini et~al., ``{Detecting gravitational waves from cosmological phase
  transitions with LISA: an
  update},''\href{http://dx.doi.org/10.1088/1475-7516/2020/03/024}{\emph{JCAP}
  {\bf 03} (2020) 024}, [\href{https://arxiv.org/abs/1910.13125}{{\tt
  1910.13125}}].

\bibitem{Athron:2023xlk}
P.~Athron, C.~Bal{\'a}zs, A.~Fowlie, L.~Morris and L.~Wu, ``{Cosmological phase
  transitions: From perturbative particle physics to gravitational
  waves},''\href{http://dx.doi.org/10.1016/j.ppnp.2023.104094}{\emph{Prog.
  Part. Nucl. Phys.} {\bf 135} (2024) 104094},
  [\href{https://arxiv.org/abs/2305.02357}{{\tt 2305.02357}}].

\bibitem{Baldes:2021vyz}
I.~Baldes, S.~Blasi, A.~Mariotti, A.~Sevrin and K.~Turbang, ``{Baryogenesis via
  relativistic bubble
  expansion},''\href{http://dx.doi.org/10.1103/PhysRevD.104.115029}{\emph{Phys.
  Rev. D} {\bf 104} (2021) 115029},
  [\href{https://arxiv.org/abs/2106.15602}{{\tt 2106.15602}}].

\bibitem{Huang:2022vkf}
P.~Huang and K.-P. Xie, ``{Leptogenesis triggered by a first-order phase
  transition},''\href{http://dx.doi.org/10.1007/JHEP09(2022)052}{\emph{JHEP}
  {\bf 09} (2022) 052}, [\href{https://arxiv.org/abs/2206.04691}{{\tt
  2206.04691}}].

\bibitem{Chun:2023ezg}
E.~J. Chun, T.~P. Dutka, T.~H. Jung, X.~Nagels and M.~Vanvlasselaer,
  ``{Bubble-assisted
  leptogenesis},''\href{http://dx.doi.org/10.1007/JHEP09(2023)164}{\emph{JHEP}
  {\bf 09} (2023) 164}, [\href{https://arxiv.org/abs/2305.10759}{{\tt
  2305.10759}}].

\bibitem{Dichtl:2023xqd}
M.~Dichtl, J.~Nava, S.~Pascoli and F.~Sala, ``{Baryogenesis and leptogenesis
  from supercooled
  confinement},''\href{http://dx.doi.org/10.1007/JHEP02(2024)059}{\emph{JHEP}
  {\bf 02} (2024) 059}, [\href{https://arxiv.org/abs/2312.09282}{{\tt
  2312.09282}}].

\bibitem{Azatov:2021irb}
A.~Azatov, M.~Vanvlasselaer and W.~Yin, ``{Baryogenesis via relativistic bubble
  walls},''\href{http://dx.doi.org/10.1007/JHEP10(2021)043}{\emph{JHEP} {\bf
  10} (2021) 043}, [\href{https://arxiv.org/abs/2106.14913}{{\tt 2106.14913}}].

\bibitem{Cataldi:2024pgt}
M.~Cataldi and B.~Shakya, ``{Leptogenesis via bubble
  collisions},''\href{http://dx.doi.org/10.1088/1475-7516/2024/11/047}{\emph{JCAP}
  {\bf 11} (2024) 047}, [\href{https://arxiv.org/abs/2407.16747}{{\tt
  2407.16747}}].

\bibitem{Pascoli:2016gkf}
S.~Pascoli, J.~Turner and Y.-L. Zhou, ``{Baryogenesis via leptonic CP-violating
  phase
  transition},''\href{http://dx.doi.org/10.1016/j.physletb.2018.03.011}{\emph{Phys.
  Lett. B} {\bf 780} (2018) 313--318},
  [\href{https://arxiv.org/abs/1609.07969}{{\tt 1609.07969}}].

\bibitem{Long:2017rdo}
A.~J. Long, A.~Tesi and L.-T. Wang, ``{Baryogenesis at a Lepton-Number-Breaking
  Phase
  Transition},''\href{http://dx.doi.org/10.1007/JHEP10(2017)095}{\emph{JHEP}
  {\bf 10} (2017) 095}, [\href{https://arxiv.org/abs/1703.04902}{{\tt
  1703.04902}}].

\bibitem{Fernandez-Martinez:2020szk}
E.~Fern{\'a}ndez-Mart{\'\i}nez, J.~L{\'o}pez-Pav{\'o}n, T.~Ota and
  S.~Rosauro-Alcaraz, ``{$\nu$ electroweak
  baryogenesis},''\href{http://dx.doi.org/10.1007/JHEP10(2020)063}{\emph{JHEP}
  {\bf 10} (2020) 063}, [\href{https://arxiv.org/abs/2007.11008}{{\tt
  2007.11008}}].

\bibitem{Chung:2012vg}
D.~J.~H. Chung, A.~J. Long and L.-T. Wang, ``{125 GeV Higgs boson and
  electroweak phase transition model
  classes},''\href{http://dx.doi.org/10.1103/PhysRevD.87.023509}{\emph{Phys.
  Rev. D} {\bf 87} (2013) 023509}, [\href{https://arxiv.org/abs/1209.1819}{{\tt
  1209.1819}}].

\bibitem{Angelescu:2018dkk}
A.~Angelescu and P.~Huang, ``{Multistep Strongly First Order Phase Transitions
  from New Fermions at the TeV
  Scale},''\href{http://dx.doi.org/10.1103/PhysRevD.99.055023}{\emph{Phys. Rev.
  D} {\bf 99} (2019) 055023}, [\href{https://arxiv.org/abs/1812.08293}{{\tt
  1812.08293}}].

\bibitem{Buen-Abad:2023hex}
M.~A. Buen-Abad, J.~H. Chang and A.~Hook, ``{Gravitational wave signatures from
  reheating},''\href{http://dx.doi.org/10.1103/PhysRevD.108.036006}{\emph{Phys.
  Rev. D} {\bf 108} (2023) 036006},
  [\href{https://arxiv.org/abs/2305.09712}{{\tt 2305.09712}}].

\bibitem{Barni:2025ced}
G.~Barni and A.~Tesi, ``{Super-heated first order phase transitions},''
  \href{https://arxiv.org/abs/2508.08362}{{\tt 2508.08362}}.

\bibitem{Ai:2024cka}
W.-Y. Ai, L.~Heurtier and T.~H. Jung, ``{Primordial black holes from an
  interrupted phase transition},'' \href{https://arxiv.org/abs/2409.02175}{{\tt
  2409.02175}}.

\bibitem{Dent:2024bhi}
J.~B. Dent, B.~Dutta and M.~Rai, ``{Imprints of early universe cosmology on
  gravitational
  waves},''\href{http://dx.doi.org/10.1007/JHEP03(2025)098}{\emph{JHEP} {\bf
  03} (2025) 098}, [\href{https://arxiv.org/abs/2411.09757}{{\tt 2411.09757}}].

\bibitem{Barni:2025mud}
G.~Barni, S.~Blasi and M.~Vanvlasselaer, ``{Inverse bubbles from broken
  supersymmetry},'' \href{https://arxiv.org/abs/2503.01951}{{\tt 2503.01951}}.

\bibitem{Wyler:1982dd}
D.~Wyler and L.~Wolfenstein, ``{Massless Neutrinos in Left-Right Symmetric
  Models},''\href{http://dx.doi.org/10.1016/0550-3213(83)90482-0}{\emph{Nucl.
  Phys. B} {\bf 218} (1983) 205--214}.

\bibitem{Mohapatra:1986aw}
R.~N. Mohapatra, ``{Mechanism for Understanding Small Neutrino Mass in
  Superstring
  Theories},''\href{http://dx.doi.org/10.1103/PhysRevLett.56.561}{\emph{Phys.
  Rev. Lett.} {\bf 56} (1986) 561--563}.

\bibitem{Mohapatra:1986bd}
R.~N. Mohapatra and J.~W.~F. Valle, ``{Neutrino Mass and Baryon Number
  Nonconservation in Superstring
  Models},''\href{http://dx.doi.org/10.1103/PhysRevD.34.1642}{\emph{Phys. Rev.
  D} {\bf 34} (1986) 1642}.

\bibitem{Peskin:1990zt}
M.~E. Peskin and T.~Takeuchi, ``{A New constraint on a strongly interacting
  Higgs
  sector},''\href{http://dx.doi.org/10.1103/PhysRevLett.65.964}{\emph{Phys.
  Rev. Lett.} {\bf 65} (1990) 964--967}.

\bibitem{Peskin:1991sw}
M.~E. Peskin and T.~Takeuchi, ``{Estimation of oblique electroweak
  corrections},''\href{http://dx.doi.org/10.1103/PhysRevD.46.381}{\emph{Phys.
  Rev. D} {\bf 46} (1992) 381--409}.

\bibitem{Kearney:2012zi}
J.~Kearney, A.~Pierce and N.~Weiner, ``{Vectorlike Fermions and Higgs
  Couplings},''\href{http://dx.doi.org/10.1103/PhysRevD.86.113005}{\emph{Phys.
  Rev. D} {\bf 86} (2012) 113005}, [\href{https://arxiv.org/abs/1207.7062}{{\tt
  1207.7062}}].

\bibitem{Joglekar:2012vc}
A.~Joglekar, P.~Schwaller and C.~E.~M. Wagner, ``{Dark Matter and Enhanced
  Higgs to Di-photon Rate from Vector-like
  Leptons},''\href{http://dx.doi.org/10.1007/JHEP12(2012)064}{\emph{JHEP} {\bf
  12} (2012) 064}, [\href{https://arxiv.org/abs/1207.4235}{{\tt 1207.4235}}].

\bibitem{Niemi:2024vzw}
L.~Niemi and T.~V.~I. Tenkanen, ``{Investigating two-loop effects for
  first-order electroweak phase
  transitions},''\href{http://dx.doi.org/10.1103/PhysRevD.111.075034}{\emph{Phys.
  Rev. D} {\bf 111} (2025) 075034},
  [\href{https://arxiv.org/abs/2408.15912}{{\tt 2408.15912}}].

\bibitem{Ramsey-Musolf:2024ykk}
M.~J. Ramsey-Musolf, T.~V.~I. Tenkanen and V.~Q. Tran, ``{Refining
  Gravitational Wave and Collider Physics Dialogue via Singlet Scalar
  Extension},'' \href{https://arxiv.org/abs/2409.17554}{{\tt 2409.17554}}.

\bibitem{Laine:2017hdk}
M.~Laine, M.~Meyer and G.~Nardini, ``{Thermal phase transition with full 2-loop
  effective
  potential},''\href{http://dx.doi.org/10.1016/j.nuclphysb.2017.04.023}{\emph{Nucl.
  Phys. B} {\bf 920} (2017) 565--600},
  [\href{https://arxiv.org/abs/1702.07479}{{\tt 1702.07479}}].

\bibitem{Funakubo:2012qc}
K.~Funakubo and E.~Senaha, ``{Two-loop effective potential, thermal
  resummation, and first-order phase transitions: Beyond the high-temperature
  expansion},''\href{http://dx.doi.org/10.1103/PhysRevD.87.054003}{\emph{Phys.
  Rev. D} {\bf 87} (2013) 054003}, [\href{https://arxiv.org/abs/1210.1737}{{\tt
  1210.1737}}].

\bibitem{Ekstedt:2022bff}
A.~Ekstedt, P.~Schicho and T.~V.~I. Tenkanen, ``{DRalgo: A package for
  effective field theory approach for thermal phase
  transitions},''\href{http://dx.doi.org/10.1016/j.cpc.2023.108725}{\emph{Comput.
  Phys. Commun.} {\bf 288} (2023) 108725},
  [\href{https://arxiv.org/abs/2205.08815}{{\tt 2205.08815}}].

\bibitem{Laine:1993ey}
M.~Laine, ``{Bubble growth as a
  detonation},''\href{http://dx.doi.org/10.1103/PhysRevD.49.3847}{\emph{Phys.
  Rev. D} {\bf 49} (1994) 3847--3853},
  [\href{https://arxiv.org/abs/hep-ph/9309242}{{\tt hep-ph/9309242}}].

\bibitem{Kurki-Suonio:1995rrv}
H.~Kurki-Suonio and M.~Laine, ``{Supersonic deflagrations in cosmological phase
  transitions},''\href{http://dx.doi.org/10.1103/PhysRevD.51.5431}{\emph{Phys.
  Rev. D} {\bf 51} (1995) 5431--5437},
  [\href{https://arxiv.org/abs/hep-ph/9501216}{{\tt hep-ph/9501216}}].

\bibitem{Espinosa:2010hh}
J.~R. Espinosa, T.~Konstandin, J.~M. No and G.~Servant, ``{Energy Budget of
  Cosmological First-order Phase
  Transitions},''\href{http://dx.doi.org/10.1088/1475-7516/2010/06/028}{\emph{JCAP}
  {\bf 06} (2010) 028}, [\href{https://arxiv.org/abs/1004.4187}{{\tt
  1004.4187}}].

\bibitem{Liu:1992tn}
B.-H. Liu, L.~D. McLerran and N.~Turok, ``{Bubble nucleation and growth at a
  baryon number producing electroweak phase
  transition},''\href{http://dx.doi.org/10.1103/PhysRevD.46.2668}{\emph{Phys.
  Rev. D} {\bf 46} (1992) 2668--2688}.

\bibitem{Moore:1995ua}
G.~D. Moore and T.~Prokopec, ``{Bubble wall velocity in a first order
  electroweak phase
  transition},''\href{http://dx.doi.org/10.1103/PhysRevLett.75.777}{\emph{Phys.
  Rev. Lett.} {\bf 75} (1995) 777--780},
  [\href{https://arxiv.org/abs/hep-ph/9503296}{{\tt hep-ph/9503296}}].

\bibitem{Moore:1995si}
G.~D. Moore and T.~Prokopec, ``{How fast can the wall move? A Study of the
  electroweak phase transition
  dynamics},''\href{http://dx.doi.org/10.1103/PhysRevD.52.7182}{\emph{Phys.
  Rev. D} {\bf 52} (1995) 7182--7204},
  [\href{https://arxiv.org/abs/hep-ph/9506475}{{\tt hep-ph/9506475}}].

\bibitem{DeCurtis:2022hlx}
S.~De~Curtis, L.~D. Rose, A.~Guiggiani, {\'A}.~G. Muyor and G.~Panico,
  ``{Bubble wall dynamics at the electroweak phase
  transition},''\href{http://dx.doi.org/10.1007/JHEP03(2022)163}{\emph{JHEP}
  {\bf 03} (2022) 163}, [\href{https://arxiv.org/abs/2201.08220}{{\tt
  2201.08220}}].

\bibitem{Laurent:2022jrs}
B.~Laurent and J.~M. Cline, ``{First principles determination of bubble wall
  velocity},''\href{http://dx.doi.org/10.1103/PhysRevD.106.023501}{\emph{Phys.
  Rev. D} {\bf 106} (2022) 023501},
  [\href{https://arxiv.org/abs/2204.13120}{{\tt 2204.13120}}].

\bibitem{Ekstedt:2024fyq}
A.~Ekstedt, O.~Gould, J.~Hirvonen, B.~Laurent, L.~Niemi, P.~Schicho et~al.,
  ``{How fast does the WallGo? A package for computing wall velocities in
  first-order phase
  transitions},''\href{http://dx.doi.org/10.1007/JHEP04(2025)101}{\emph{JHEP}
  {\bf 04} (2025) 101}, [\href{https://arxiv.org/abs/2411.04970}{{\tt
  2411.04970}}].

\bibitem{Branchina:2025jou}
C.~Branchina, A.~Conaci, S.~De~Curtis and L.~Delle~Rose, ``{Electroweak Phase
  Transition and Bubble Wall Velocity in Local Thermal Equilibrium},''
  \href{https://arxiv.org/abs/2504.21213}{{\tt 2504.21213}}.

\bibitem{Bodeker:2009qy}
D.~Bodeker and G.~D. Moore, ``{Can electroweak bubble walls run
  away?},''\href{http://dx.doi.org/10.1088/1475-7516/2009/05/009}{\emph{JCAP}
  {\bf 05} (2009) 009}, [\href{https://arxiv.org/abs/0903.4099}{{\tt
  0903.4099}}].

\bibitem{Bodeker:2017cim}
D.~Bodeker and G.~D. Moore, ``{Electroweak Bubble Wall Speed
  Limit},''\href{http://dx.doi.org/10.1088/1475-7516/2017/05/025}{\emph{JCAP}
  {\bf 05} (2017) 025}, [\href{https://arxiv.org/abs/1703.08215}{{\tt
  1703.08215}}].

\bibitem{Azatov:2020ufh}
A.~Azatov and M.~Vanvlasselaer, ``{Bubble wall velocity: heavy physics
  effects},''\href{http://dx.doi.org/10.1088/1475-7516/2021/01/058}{\emph{JCAP}
  {\bf 01} (2021) 058}, [\href{https://arxiv.org/abs/2010.02590}{{\tt
  2010.02590}}].

\bibitem{Hoche:2020ysm}
S.~H{\"o}che, J.~Kozaczuk, A.~J. Long, J.~Turner and Y.~Wang, ``{Towards an
  all-orders calculation of the electroweak bubble wall
  velocity},''\href{http://dx.doi.org/10.1088/1475-7516/2021/03/009}{\emph{JCAP}
  {\bf 03} (2021) 009}, [\href{https://arxiv.org/abs/2007.10343}{{\tt
  2007.10343}}].

\bibitem{Gouttenoire:2021kjv}
Y.~Gouttenoire, R.~Jinno and F.~Sala, ``{Friction pressure on relativistic
  bubble walls},''\href{http://dx.doi.org/10.1007/JHEP05(2022)004}{\emph{JHEP}
  {\bf 05} (2022) 004}, [\href{https://arxiv.org/abs/2112.07686}{{\tt
  2112.07686}}].

\bibitem{Ai:2023suz}
W.-Y. Ai, ``{Logarithmically divergent friction on ultrarelativistic bubble
  walls},''\href{http://dx.doi.org/10.1088/1475-7516/2023/10/052}{\emph{JCAP}
  {\bf 10} (2023) 052}, [\href{https://arxiv.org/abs/2308.10679}{{\tt
  2308.10679}}].

\bibitem{Azatov:2023xem}
A.~Azatov, G.~Barni, R.~Petrossian-Byrne and M.~Vanvlasselaer, ``{Quantisation
  across bubble walls and
  friction},''\href{http://dx.doi.org/10.1007/JHEP05(2024)294}{\emph{JHEP} {\bf
  05} (2024) 294}, [\href{https://arxiv.org/abs/2310.06972}{{\tt 2310.06972}}].

\bibitem{Konstandin:2010dm}
T.~Konstandin and J.~M. No, ``{Hydrodynamic obstruction to bubble
  expansion},''\href{http://dx.doi.org/10.1088/1475-7516/2011/02/008}{\emph{JCAP}
  {\bf 02} (2011) 008}, [\href{https://arxiv.org/abs/1011.3735}{{\tt
  1011.3735}}].

\bibitem{Cline:2021iff}
J.~M. Cline, A.~Friedlander, D.-M. He, K.~Kainulainen, B.~Laurent and
  D.~Tucker-Smith, ``{Baryogenesis and gravity waves from a UV-completed
  electroweak phase
  transition},''\href{http://dx.doi.org/10.1103/PhysRevD.103.123529}{\emph{Phys.
  Rev. D} {\bf 103} (2021) 123529},
  [\href{https://arxiv.org/abs/2102.12490}{{\tt 2102.12490}}].

\bibitem{Ai:2023see}
W.-Y. Ai, B.~Laurent and J.~van~de Vis, ``{Model-independent bubble wall
  velocities in local thermal
  equilibrium},''\href{http://dx.doi.org/10.1088/1475-7516/2023/07/002}{\emph{JCAP}
  {\bf 07} (2023) 002}, [\href{https://arxiv.org/abs/2303.10171}{{\tt
  2303.10171}}].

\bibitem{Ai:2024shx}
W.-Y. Ai, X.~Nagels and M.~Vanvlasselaer, ``{Criterion for ultra-fast bubble
  walls: the impact of hydrodynamic
  obstruction},''\href{http://dx.doi.org/10.1088/1475-7516/2024/03/037}{\emph{JCAP}
  {\bf 03} (2024) 037}, [\href{https://arxiv.org/abs/2401.05911}{{\tt
  2401.05911}}].

\bibitem{Azatov:2024auq}
A.~Azatov, G.~Barni and R.~Petrossian-Byrne, ``{NLO friction in symmetry
  restoring phase
  transitions},''\href{http://dx.doi.org/10.1007/JHEP12(2024)056}{\emph{JHEP}
  {\bf 12} (2024) 056}, [\href{https://arxiv.org/abs/2405.19447}{{\tt
  2405.19447}}].

\bibitem{Malinsky:2009df}
M.~Malinsky, T.~Ohlsson, Z.-z. Xing and H.~Zhang, ``{Non-unitary neutrino
  mixing and CP violation in the minimal inverse seesaw
  model},''\href{http://dx.doi.org/10.1016/j.physletb.2009.07.038}{\emph{Phys.
  Lett. B} {\bf 679} (2009) 242--248},
  [\href{https://arxiv.org/abs/0905.2889}{{\tt 0905.2889}}].

\bibitem{Abada:2014vea}
A.~Abada and M.~Lucente, ``{Looking for the minimal inverse seesaw
  realisation},''\href{http://dx.doi.org/10.1016/j.nuclphysb.2014.06.003}{\emph{Nucl.
  Phys. B} {\bf 885} (2014) 651--678},
  [\href{https://arxiv.org/abs/1401.1507}{{\tt 1401.1507}}].

\bibitem{Casas:2001sr}
J.~A. Casas and A.~Ibarra, ``{Oscillating neutrinos and $\mu \to e,
  \gamma$},''\href{http://dx.doi.org/10.1016/S0550-3213(01)00475-8}{\emph{Nucl.
  Phys. B} {\bf 618} (2001) 171--204},
  [\href{https://arxiv.org/abs/hep-ph/0103065}{{\tt hep-ph/0103065}}].

\bibitem{Dolan:2018qpy}
M.~J. Dolan, T.~P. Dutka and R.~R. Volkas, ``{Dirac-Phase Thermal Leptogenesis
  in the extended Type-I Seesaw
  Model},''\href{http://dx.doi.org/10.1088/1475-7516/2018/06/012}{\emph{JCAP}
  {\bf 06} (2018) 012}, [\href{https://arxiv.org/abs/1802.08373}{{\tt
  1802.08373}}].

\bibitem{Capozzi:2021fjo}
F.~Capozzi, E.~Di~Valentino, E.~Lisi, A.~Marrone, A.~Melchiorri and A.~Palazzo,
  ``{Unfinished fabric of the three neutrino
  paradigm},''\href{http://dx.doi.org/10.1103/PhysRevD.104.083031}{\emph{Phys.
  Rev. D} {\bf 104} (2021) 083031},
  [\href{https://arxiv.org/abs/2107.00532}{{\tt 2107.00532}}].

\bibitem{Azatov:2021ifm}
A.~Azatov, M.~Vanvlasselaer and W.~Yin, ``{Dark Matter production from
  relativistic bubble
  walls},''\href{http://dx.doi.org/10.1007/JHEP03(2021)288}{\emph{JHEP} {\bf
  03} (2021) 288}, [\href{https://arxiv.org/abs/2101.05721}{{\tt 2101.05721}}].

\bibitem{Baldes:2022oev}
I.~Baldes, Y.~Gouttenoire and F.~Sala, ``{Hot and heavy dark matter from a weak
  scale phase
  transition},''\href{http://dx.doi.org/10.21468/SciPostPhys.14.3.033}{\emph{SciPost
  Phys.} {\bf 14} (2023) 033}, [\href{https://arxiv.org/abs/2207.05096}{{\tt
  2207.05096}}].

\bibitem{Azatov:2024crd}
A.~Azatov, X.~Nagels, M.~Vanvlasselaer and W.~Yin, ``{Populating secluded dark
  sector with ultra-relativistic
  bubbles},''\href{http://dx.doi.org/10.1007/JHEP11(2024)129}{\emph{JHEP} {\bf
  11} (2024) 129}, [\href{https://arxiv.org/abs/2406.12554}{{\tt 2406.12554}}].

\bibitem{Ai:2024ikj}
W.-Y. Ai, M.~Fairbairn, K.~Mimasu and T.~You, ``{Non-thermal production of
  heavy vector dark matter from relativistic bubble
  walls},''\href{http://dx.doi.org/10.1007/JHEP05(2025)225}{\emph{JHEP} {\bf
  05} (2025) 225}, [\href{https://arxiv.org/abs/2406.20051}{{\tt 2406.20051}}].

\bibitem{Ai:2025bjw}
W.-Y. Ai, M.~Carosi, B.~Garbrecht, C.~Tamarit and M.~Vanvlasselaer, ``{Bubble
  wall dynamics from nonequilibrium quantum field
  theory},''\href{http://dx.doi.org/10.1007/JHEP08(2025)077}{\emph{JHEP} {\bf
  08} (2025) 077}, [\href{https://arxiv.org/abs/2504.13725}{{\tt 2504.13725}}].

\bibitem{Ramsey-Musolf:2025jyk}
M.~J. Ramsey-Musolf and J.~Zhu, ``{Bubble wall velocity from Kadanoff-Baym
  equations: fluid dynamics and microscopic interactions},''
  \href{https://arxiv.org/abs/2504.13724}{{\tt 2504.13724}}.

\bibitem{ALEPH:2005ab}
{\scshape ALEPH, DELPHI, L3, OPAL, SLD, LEP Electroweak Working Group, SLD
  Electroweak Group, SLD Heavy Flavour Group} collaboration, S.~Schael et~al.,
  ``{Precision electroweak measurements on the $Z$
  resonance},''\href{http://dx.doi.org/10.1016/j.physrep.2005.12.006}{\emph{Phys.
  Rept.} {\bf 427} (2006) 257--454},
  [\href{https://arxiv.org/abs/hep-ex/0509008}{{\tt hep-ex/0509008}}].

\bibitem{Pilaftsis:1997jf}
A.~Pilaftsis, ``{CP violation and baryogenesis due to heavy Majorana
  neutrinos},''\href{http://dx.doi.org/10.1103/PhysRevD.56.5431}{\emph{Phys.
  Rev. D} {\bf 56} (1997) 5431--5451},
  [\href{https://arxiv.org/abs/hep-ph/9707235}{{\tt hep-ph/9707235}}].

\bibitem{Pilaftsis:2003gt}
A.~Pilaftsis and T.~E.~J. Underwood, ``{Resonant
  leptogenesis},''\href{http://dx.doi.org/10.1016/j.nuclphysb.2004.05.029}{\emph{Nucl.
  Phys. B} {\bf 692} (2004) 303--345},
  [\href{https://arxiv.org/abs/hep-ph/0309342}{{\tt hep-ph/0309342}}].

\bibitem{Adhikary:2014qba}
B.~Adhikary, M.~Chakraborty and A.~Ghosal, ``{Flavored leptogenesis with
  quasidegenerate neutrinos in a broken cyclic symmetric
  model},''\href{http://dx.doi.org/10.1103/PhysRevD.93.113001}{\emph{Phys. Rev.
  D} {\bf 93} (2016) 113001}, [\href{https://arxiv.org/abs/1407.6173}{{\tt
  1407.6173}}].

\bibitem{Branco:2011zb}
G.~C. Branco, R.~G. Felipe and F.~R. Joaquim, ``{Leptonic CP
  Violation},''\href{http://dx.doi.org/10.1103/RevModPhys.84.515}{\emph{Rev.
  Mod. Phys.} {\bf 84} (2012) 515--565},
  [\href{https://arxiv.org/abs/1111.5332}{{\tt 1111.5332}}].

\bibitem{Li:2021tlv}
S.-P. Li, X.-Q. Li, X.-S. Yan and Y.-D. Yang, ``{Baryogenesis from hierarchical
  Dirac
  neutrinos},''\href{http://dx.doi.org/10.1103/PhysRevD.104.115014}{\emph{Phys.
  Rev. D} {\bf 104} (2021) 115014},
  [\href{https://arxiv.org/abs/2105.01317}{{\tt 2105.01317}}].

\bibitem{Mukherjee:2023nyi}
A.~Mukherjee and A.~K. Saha, ``{Rescuing leptogenesis parameter space of
  inverse
  seesaw},''\href{http://dx.doi.org/10.1016/j.physletb.2024.138474}{\emph{Phys.
  Lett. B} {\bf 849} (2024) 138474},
  [\href{https://arxiv.org/abs/2307.14405}{{\tt 2307.14405}}].

\bibitem{Shao:2025lym}
Y.~Shao and Z.-h. Zhao, ``{Rescuing leptogenesis in inverse seesaw models with
  the help of non-Abelian flavor symmetries},''
  \href{https://arxiv.org/abs/2504.19697}{{\tt 2504.19697}}.

\bibitem{Deppisch:2010fr}
F.~F. Deppisch and A.~Pilaftsis, ``{Lepton Flavour Violation and theta(13) in
  Minimal Resonant
  Leptogenesis},''\href{http://dx.doi.org/10.1103/PhysRevD.83.076007}{\emph{Phys.
  Rev. D} {\bf 83} (2011) 076007}, [\href{https://arxiv.org/abs/1012.1834}{{\tt
  1012.1834}}].

\bibitem{Dasgupta:2022isg}
A.~Dasgupta, P.~S.~B. Dev, A.~Ghoshal and A.~Mazumdar, ``{Gravitational wave
  pathway to testable
  leptogenesis},''\href{http://dx.doi.org/10.1103/PhysRevD.106.075027}{\emph{Phys.
  Rev. D} {\bf 106} (2022) 075027},
  [\href{https://arxiv.org/abs/2206.07032}{{\tt 2206.07032}}].

\bibitem{ATLAS:2024mrr}
{\scshape ATLAS} collaboration, G.~Aad et~al., ``{Search for vector-like
  leptons coupling to first- and second-generation Standard Model leptons in pp
  collisions at $ \sqrt{s} $ = 13 TeV with the ATLAS
  detector},''\href{http://dx.doi.org/10.1007/JHEP05(2025)075}{\emph{JHEP} {\bf
  05} (2025) 075}, [\href{https://arxiv.org/abs/2411.07143}{{\tt 2411.07143}}].

\bibitem{ATLAS:2025wgc}
{\scshape ATLAS} collaboration, G.~Aad et~al., ``{Search for electroweak
  production of vector-like leptons in $\tau$-lepton and $b$-jet final states
  in $pp$ collisions at $\sqrt{s}$ = 13 TeV with the ATLAS detector},''
  \href{https://arxiv.org/abs/2503.22581}{{\tt 2503.22581}}.

\bibitem{Liss:2013hbb}
{\scshape ATLAS} collaboration, A.~Liss and J.~Nielsen, ``{Physics at a
  High-Luminosity LHC with ATLAS},''
  \href{https://arxiv.org/abs/1307.7292}{{\tt 1307.7292}}.

\bibitem{CMS:2013xfa}
{\scshape CMS} collaboration, ``{Projected Performance of an Upgraded CMS
  Detector at the LHC and HL-LHC: Contribution to the Snowmass Process},'' in
  \emph{{Snowmass 2013}: {Snowmass on the Mississippi}}, 7, 2013.
\newblock \href{https://arxiv.org/abs/1307.7135}{{\tt 1307.7135}}.

\bibitem{ATL-PHYS-PUB-2014-016}
``{Projections for measurements of Higgs boson signal strengths and coupling
  parameters with the ATLAS detector at a HL-LHC},'' tech. rep., CERN, Geneva,
  2014.
\newblock 10.17181/CERN.B5WP.VPT7.

\bibitem{Crowder:2005nr}
J.~Crowder and N.~J. Cornish, ``{Beyond LISA: Exploring future gravitational
  wave
  missions},''\href{http://dx.doi.org/10.1103/PhysRevD.72.083005}{\emph{Phys.
  Rev. D} {\bf 72} (2005) 083005},
  [\href{https://arxiv.org/abs/gr-qc/0506015}{{\tt gr-qc/0506015}}].

\bibitem{Guo:2023jkz}
Q.~Guo, L.~Gao, Y.~Mao and Q.~Li, ``{Vector-like lepton searches at a muon
  collider in the context of the 4321
  model*},''\href{http://dx.doi.org/10.1088/1674-1137/ace5a7}{\emph{Chin. Phys.
  C} {\bf 47} (2023) 103106}, [\href{https://arxiv.org/abs/2304.01885}{{\tt
  2304.01885}}].

\bibitem{Liu:2021akf}
W.~Liu, K.-P. Xie and Z.~Yi, ``{Testing leptogenesis at the LHC and future muon
  colliders: A Z'
  scenario},''\href{http://dx.doi.org/10.1103/PhysRevD.105.095034}{\emph{Phys.
  Rev. D} {\bf 105} (2022) 095034},
  [\href{https://arxiv.org/abs/2109.15087}{{\tt 2109.15087}}].

\bibitem{Coleman:1973jx}
S.~R. Coleman and E.~J. Weinberg, ``{Radiative Corrections as the Origin of
  Spontaneous Symmetry
  Breaking},''\href{http://dx.doi.org/10.1103/PhysRevD.7.1888}{\emph{Phys. Rev.
  D} {\bf 7} (1973) 1888--1910}.

\bibitem{Linde:1981zj}
A.~D. Linde, ``{Decay of the False Vacuum at Finite
  Temperature},''\href{http://dx.doi.org/10.1016/0550-3213(83)90072-X}{\emph{Nucl.
  Phys. B} {\bf 216} (1983) 421}.

\bibitem{Quiros:1999jp}
M.~Quiros, ``{Finite temperature field theory and phase transitions},'' in
  \emph{{ICTP Summer School in High-Energy Physics and Cosmology}},
  pp.~187--259, 1, 1999.
\newblock \href{https://arxiv.org/abs/hep-ph/9901312}{{\tt hep-ph/9901312}}.

\end{thebibliography}\endgroup

\end{document}